\documentclass[12pt]{article}

\usepackage{amsmath,amssymb,cite,bm}
\usepackage{graphicx}

\makeatletter
 
  \@addtoreset{equation}{section}
 \makeatother

\newcommand{\n}{\nonumber \\}
\newcommand{\Tr}{\mathrm{Tr}}

\newcommand{\be}{\begin{equation}}
\newcommand{\ee}{\end{equation}}
\newcommand{\bea}{\begin{eqnarray}}
\newcommand{\eea}{\end{eqnarray}}
\newcommand{\beann}{\begin{eqnarray*}}
\newcommand{\eeann}{\end{eqnarray*}}

\newcommand{\ba}{\begin{array}}
\newcommand{\ea}{\end{array}}

\allowdisplaybreaks

\begin{document}

\setlength{\oddsidemargin}{0cm}
\setlength{\baselineskip}{7mm}

\begin{titlepage}
\renewcommand{\thefootnote}{\fnsymbol{footnote}}
\begin{normalsize}
\begin{flushright}
\begin{tabular}{l}
YITP-13-123 \\
KUNS-2470 
\end{tabular}
\end{flushright}
  \end{normalsize}

~~\\

\vspace*{0cm}
    \begin{Large}
%    \begin{bf}
       \begin{center}
         {Emergent bubbling geometries in the plane wave matrix model}
       \end{center}
%    \end{bf}   
    \end{Large}
\vspace{0.7cm}

\begin{center}
Yuhma A{\sc sano}$^{1)}$\footnote
            {
e-mail address : 
yuhma@gauge.scphys.kyoto-u.ac.jp }, 
Goro I{\sc shiki}$^{1),2)}$\footnote
            {
e-mail address : 
ishiki@yukawa.kyoto-u.ac.jp}, 
Takashi O{\sc kada}$^{1),2),3)}$\footnote
            {
e-mail address : 
okada@yukawa.kyoto-u.ac.jp }
    {\sc and}
Shinji S{\sc himasaki}$^{1)}$\footnote
           {
e-mail address : 
shinji@gauge.scphys.kyoto-u.ac.jp }\\
      
\vspace{0.7cm}
                    
     $^{1)}$ {\it Department of Physics, Kyoto University}\\
               {\it Kyoto, 606-8502, Japan}\\

     $^{2)}$ {\it Yukawa Institute for Theoretical Physics, Kyoto University}\\
               {\it Kyoto, 606-8502, Japan}\\

     $^{3)}$ {\it Kavli Institute for Theoretical Physics, UCSB}\\
               {\it Santa Barbara CA 93106}\\

\end{center}

\vspace{0.7cm}

\begin{abstract}
\noindent
The gravity dual geometry of the plane wave matrix model is 
given by the bubbling geometry in the type IIA supergravity, 
which is described by an axially symmetric electrostatic system.
We study a quarter BPS sector of the plane wave matrix model 
in terms of the localization method and show that this sector 
can be mapped to a one-dimensional interacting Fermi gas system.
We find that the mean-field density of the Fermi gas can be identified with 
the charge density in the electrostatic system in the gravity side.
We also find that the scaling limits in which the dual geometry 
reduces to the D2-brane or NS5-brane geometry are given as the 
free limit or the strongly coupled limit of the Fermi gas system,
respectively. We reproduce the radii of $S^5$'s in these geometries 
by solving the Fermi gas model in the corresponding limits.

%This result agrees with the gravity dual picture where 
%the same scaling limit had already been discovered.
%
%This double scaling limit was firstly discovered 
%on the gravity dual side of the plane wave matrix.
%Hence this result also provides a non-trivial check of the gauge/gravity 
%correspondence.

\end{abstract}
\vfill

\end{titlepage}
\vfil\eject

\setcounter{footnote}{0}

%\tableofcontents

%%%%%%%%%%%%%%%%%%%%%%%%%%%%%%%%%%%%%%%%%%%%%%%%%%%%%%%%%%%%%
\section{Introduction}
%%%%%%%%%%%%%%%%%%%%%%%%%%%%%%%%%%%%%%%%%%%%%%%%%%%%%%%%%%%%%
The gauge/string duality is a conjectured equivalence
between strongly coupled gauge field theories and 
weakly coupled perturbative string theories \cite{Maldacena, GKP, Witten}.
One of the most important problems in understanding this duality
is how the space-time geometry described in the string theory 
emerges in the framework of the corresponding gauge theory.
If the duality is true, the background space-time in the string theory 
should emerge in the strong coupling region of the 
gauge theories, although it may not be apparent in 
the weak coupling region.
See \cite{Berenstein:2005aa,Berenstein:2008eg,Steinacker:2010rh,
Kim:2011cr,Yang:2013aia} for recent developments on 
the emergent space-time.

In this paper, we analyze a one-dimensional gauge theory 
in the strong coupling region and study the emergent 
phenomena of geometries.
The theory we consider is a matrix quantum mechanics 
called the plane wave matrix model (PWMM), 
which was proposed as a fundamental 
formulation of the M-theory on the pp-wave background 
in the light-cone frame \cite{BMN}. 
PWMM has $SU(2|4)$ symmetry, which consists of $R\times SO(3)\times SO(6)$
bosonic symmetry and 16 supersymmetries.
PWMM is a mass deformation of the BFSS matrix model \cite{Banks:1996vh} 
 and has many discrete vacua
given by fuzzy spheres that are labeled 
by representations of the $SU(2)$ Lie algebra.
%whose dimension is equal to the matrix size of PWMM.
%Since every representation ${\cal R}$ has 
%an unique irreducible decomposition, one can also
%label the vacuum by a set of two positive integers, 
%$\{(N_2^{(s)},N_5^{(s)})| s=1,2,\cdots,\Lambda\}$, where 
%$N_2^{(s)}$ and $N_5^{(s)}$ correspond to the multiplicity and the 
%dimension of
%each irreducible representation that appears in the decomposition, 
%respectively, and $\Lambda$ is the number of different 
%irreducible representations. 
In this paper, we consider the case where the representation 
is given by a direct sum of the same irreducible representations.
In this case, the vacua are labeled by two integers $(N_2,N_5)$, where 
$N_5$ is the dimension of the irreducible representation and 
$N_2$ is the multiplicity.

For the theory around each vacuum of PWMM,
a corresponding gravity dual geometry in the type IIA 
superstring theory was constructed in \cite{LLM,LM} 
(also studied in \cite{Lin:2004kw} in the Polchinski-Strassler approximation). The geometry is 
called the bubbling geometry and characterized by fermionic droplets on a 
certain two-dimensional subspace of the space-time, which 
define a boundary condition of the solution.
By a simple change of variables, 
the geometry can be equivalently characterized by a three-dimensional
axially symmetric electrostatic system with some conducting disks 
(see the next section).
The geometry locally has a topology of 
$R\times S^2 \times S^5 \times{\cal M}_{e}$, where ${\cal M}_{e} \sim R^2$ 
corresponds to the space on which 
the electrostatic system is defined.
For theories around the above mentioned vacua 
labeled by two integers $(N_2,N_5)$,
the gravity dual geometries were studied in detail in \cite{Ling:2006up}.
The integers, $N_2$ and $N_5$, are interpreted as the 
D2-brane and NS5-brane charges
in the dual geometries, respectively \cite{Maldacena:2002rb}.

In order to see the emergence of this geometry in PWMM, 
we consider a complex scalar field $\phi(t) \sim X_4(t)+
i(X_{9}(t)\sin t + X_{10}(t)\cos t )$, where $X_4$ is one of $SO(3)$ 
scalar fields, $X_{9,10}$ are $SO(6)$ scalars and $t$ is the time coordinate.
We consider a quarter BPS sector of PWMM which consists of correlators
of only $\phi$'s.
Since $\phi$ has two real degrees of freedom, it should describe 
a two-dimensional surface on the gravity dual geometry.
Let ${\cal M}_{\phi}$ be a two-dimensional subspace in the bubbling geometry
described by $\phi$.
For a fixed $t$, insertions of the field $\phi(t)$ 
in the path integral break
the original $R\times SO(3) \times SO(6)$ symmetry to 
$SO(2) \times SO(5)$, which is the symmetry of $S^2\times S^5$ with 
a marked point. 
%It also breaks the supersymmetries except
%four which leave $\phi$ invariant.
Hence, ${\cal M}_{\phi}$ is expected to be fibered on the marked point on
$S^2\times S^5$. 
From the original symmetry, however, 
${\cal M}_{\phi}$ should exists everywhere on $R\times S^2 \times S^5$, 
so that the topology of the total space should be locally given by
%$\frac{R\times SO(3) \times SO(6)}{SO(2)\times SO(5)} \times {\cal M}_{\phi}
%\sim 
$R\times S^2 \times S^5 \times{\cal M}_{\phi}$.
%(In other words, the definition of 
%$\phi$ picks up a point on $R\times S^2 \times S^5$ in 
%the dual gravity picutre and ${\cal M}_{\phi}$ is defined on that point.
%From the original symmetry, ${\cal M}_{\phi}$ should exists everywhere
%on $R\times S^2 \times S^5$.)
Thus, the subspace ${\cal M}_{\phi}$ can naturally be 
identified with the space ${\cal M}_{e}$ of 
the electrostatic problem.

Recently, the localization method, which makes exact 
computations possible for some supersymmetric 
operators \cite{Nekrasov:2002qd,Pestun:2007rz}, 
was applied to the above sector of $\phi$
\cite{Asano:2012zt}.
In this paper, using this result, we first show that
this sector can be mapped to a one-dimensional 
interacting Fermi gas system. 
We then find that in a strong coupling region,
the mean-field density of the Fermi particles satisfies 
the same integral equation as the charge density 
which appears in the electrostatic system in the gravity side.
Thus we identify the mean-field density with the charge density.
Since the charge density ultimately determines the gravity dual solution,
this identification makes it possible to reconstruct the 
bubbling geometry based on the gauge theory.
%In this sense, our result gives an example of the emergent 
%geometries in gauge theories.

This situation is very similar to the case of the bubbling geometries
in the type IIB supergravity which has $R\times SO(4)\times SO(4)$
symmetry and 16 supersymmetries \cite{LLM}. 
These geometries correspond to a sector of the half BPS
operators in the ${\cal N}=4$ Super Yang-Mills theory (SYM) on $R\times S^3$ 
that consists of zero modes on $S^3$ 
of a complex scalar field $Z=\Phi_4+i\Phi_5$.
This sector can also be mapped to a one-dimensional fermionic system 
and its phase space density can be identified with the
the droplets in the gravity side
\cite{Berenstein:2004kk,Takayama:2005yq}.

In the case of ${\cal N}=4$ SYM, the sector of $Z$ is protected
by the non-renormalization theorem, so it does not
depend on the coupling constant \cite{Bianchi:1999ie,Eden:1999kw}. 
%The rank of the gauge group $N$
%is usually sent to infinity to suppress the quantum effects of strings,
%so that the only tunable parameter is the power 
%$J$ of the operators, ${\rm tr}Z^J$. 
In our case, however, the sector we consider in this paper 
depends on various tunable parameters such as the coupling 
constant and the parameters $(N_2,N_5)$ of the vacua.
Hence we can also consider some scaling limits for these parameters.
The gravity dual of PWMM has two interesting scaling limits which 
we call the D2-brane limit and the NS5-brane limit in this paper
\cite{LM, Ling:2006up}.
In the D2-brane limit, 
the NS5-brane charges decouple and 
only the D2-brane charge is left in the geometry. 
The geometry asymptotically becomes the D2-brane solution. 
In the same manner, the NS5-brane limit sends the geometry 
to the NS5-brane solution.
In this paper, We show that 
the D2-brane and the NS5-brane limits are realized on the gauge theory side
as the free and strongly coupled limits of the Fermi gas system, 
respectively.
In these limits, we solve the Fermi gas system 
at the planar level and reproduce 
the radii of the $S^5$'s in the gravity side.
In particular, the $S^5$ corresponds to the spatial worldvolume of 
fivebranes in the case of the NS5-brane limit and
its radius has been known to be proportional to $\lambda^{1/4}$ in the string unit, 
where $\lambda$ is the 't Hooft coupling in 
PWMM\cite{Maldacena:2002rb}. 
Our gauge theory result reproduces this behavior and hence
gives a strong evidence for the description of 
fivebranes in PWMM proposed in 
\cite{Maldacena:2002rb}\footnote{See \cite{Aharony:1997th,Witten:1997yu,ArkaniHamed:2001ie,Douglas:2010iu} for
various descriptions of fivebranes.}.

This paper is organized as follows. 
In section~\ref{Plane wave matrix model}, 
we review PWMM and the 
result of the localization \cite{Asano:2012zt}. 
In section~\ref{Gravity dual of PWMM}, 
we review the gravity dual of PWMM.
In section~\ref{Emergent bubbling geometry}, 
we first map the matrix integral obtained 
through the localization to an interacting Fermi gas system.
Then we identify the mean-field density with the charge density.
We also perform various consistency checks of this identification.
Section~\ref{Summary and discussion}
is devoted to summary and discussion.

%%%%%%%%%%%%%%%%%%%%%%%%%%%%%%%%%%%%%%%%%%%%%%%%%%%%%%%%%%%%%
\section{Plane wave matrix model}
\label{Plane wave matrix model}
%%%%%%%%%%%%%%%%%%%%%%%%%%%%%%%%%%%%%%%%%%%%%%%%%%%%%%%%%%%%%
In this section, we review PWMM \cite{BMN} and the result 
of the localization obtained in \cite{Asano:2012zt}.
We use the same notation as in \cite{Asano:2012zt}. 
The action of PWMM is written in terms of
the ten-dimensional notation as,
\begin{align}
S&=\frac{1}{g^2}\int d\tau  \Tr\Bigl(
\frac{1}{4}F_{MN}F^{MN}
+\frac{m^2}{8}X_mX^m
+\frac{i}{2}\Psi \Gamma^M D_M \Psi
\Bigr),
\label{action of PWMM}
\end{align}
where the time direction is assumed to be the Euclidean signature and
\begin{align}
F_{1M}&=D_1X_M=\partial_1X_M-i[X_1,X_M] \quad (M\neq 1),\n
F_{a'b'}&=m \varepsilon_{a'b'c'}X_{c'}-i[X_{a'},X_{b'}], \quad
F_{a'm}=D_{a'}X_m=-i[X_{a'},X_m], \quad
F_{mn}=-i[X_m,X_n], \n
D_1\Psi&=\partial_1\Psi-i[X_1,\Psi], \quad
D_{a'}\Psi=\frac{m}{8}\varepsilon_{a'b'c'}\Gamma^{b'c'}\Psi-i[X_{a'},\Psi], \quad
D_m\Psi=-i[X_m,\Psi].
\label{F in PWMM}
\end{align}
The range of indices are $M,N=1,\cdots,10$, 
$a',b'=2,3,4$ and $m,n=5,\cdots,10$.
$X_1$ is the one-dimensional gauge field, 
$X_{a'}$ and $X_m$ are $SO(3)$ and $SO(6)$ scalars and 
$\Psi$ is a fermionic field with 16 components. 
We put $m =2$ for the mass parameter in the following. 
The $m$ dependence 
can be recovered anytime by the dimensional analysis.

The vacuum of PWMM is given by the fuzzy sphere, namely,
\begin{align}
X_a=-2 L_a, \;\; (a=2,3,4)
\label{fuzzy sphere}
\end{align}
and all the other matrices are zero.
Here $L_a$ are representation matrices of $SU(2)$ generators.
For any representation of $SU(2)$, (\ref{fuzzy sphere}) gives
a classical vacuum of PWMM which preserves 16 supersymmetries.
The representation of $L_a$ is reducible in general 
and it can be decomposed as
\begin{align}
L_a = \bigoplus_{s=1}^{\Lambda} ({\bf 1}_{N_2^{(s)}}\otimes 
L_a^{[N_5^{(s)}]}),
\label{irr decom}
\end{align}
where $L_a^{[N]}$ are $SU(2)$ generators in the $N$ dimensional irreducible
representation. $N_2^{(s)}$ denote the multiplicities of 
the irreducible representations and $\sum_s N_2^{(s)}N_5^{(s)}$ must 
be equal to the matrix size in PWMM. 
The notation for $N_2^{(s)}$ and $N_5^{(s)}$ 
indicates that they correspond to membrane and 5-brane charges 
in M-theory, respectively \cite{Maldacena:2002rb}.

In order to define the path integral of PWMM,
one has to specify the boundary conditions at $\tau \rightarrow \pm \infty$.
To study the theory around a fixed vacuum of PWMM, 
the appropriate boundary condition is such that all 
fields approach to the vacuum configuration at the both infinities. 
When the multiplicities are sufficiently large compared to 
the other parameters, 
the instanton effects\footnote{See 
\cite{Bachas:2000dx,Yee:2003ge, Lin:2006tr} 
for instanton solutions in PWMM.}
can be ignored, so that the path integral with this boundary condition
defines the theory around the fixed vacuum.

Let us define a complex scalar field,
\begin{align}
\phi (\tau) = 2(-X_4(\tau)+\sinh \tau X_9(\tau)+ i\cosh \tau X_{10}(\tau)).
\label{phi}
\end{align}
When $\tau$ is Wick-rotated to the Lorentzian signature, 
the real and the imaginary parts of $\phi$ are given by 
a $SO(3)$ scalar and $SO(6)$ scalars, respectively, 
as introduced in the previous section.
On the other hand, in another 
Lorentzian signature where the $X_{10}$ direction is Wick-rotated as 
in \cite{Pestun:2007rz}, 
four supersymmetries which leave $\phi$ invariant were constructed 
\cite{Asano:2012zt}. 
%These supersymmetry transformations are denoted by $Q$ below.
See appendix~\ref{Off-shell supersymmetry in PWMM}
for these supersymmetries.

For the theory around each vacuum of PWMM, 
one can compute the correlators made of $\phi $'s by Wick-rotating $X_{10}$
and applying the localization with the above mentioned boundary 
conditions \cite{Asano:2012zt}\footnote{See 
\cite{Sugishita:2013jca,Honda:2013uca,Hori:2013ika} 
for localization computations for the case with a boundary.}. 
%In the localization, one first constructs a functional $V$ 
%and adds a term $-u QV$ to the action, where $u$ is a deformation 
%parameter. The functional $V$ should satisfy two conditions:
%the positive definiteness of the bosonic part of $QV$ and $Q^2V$=0. 
%Then from the $Q$ invariance of the action and $\phi$,
%it follows that the deformed partition function does not depend on $u$. 
%Thus by taking $u \rightarrow \infty$, 
%expectation values of correlators made only of $\phi $'s 
%are dominated by saddle point configurations of $QV$, which 
%are parameterized by some constant matrices in our case. 
%Thus, the computation of the correlators 
%is reduced to a finite dimensional matrix integral.
%
The result of the localization in \cite{Asano:2012zt} 
is summarized below. The following equality holds,
\begin{align}
\langle \prod_{a} {\rm Tr}f_a(\phi (\tau_a)) \rangle
=\langle \prod_{a} {\rm Tr}f_a(4L_4+2iM) \rangle_{MM},
\label{result}
\end{align}
where $f_{a}$ are arbitrary smooth functions. On the left-hand side
of (\ref{result}), the expectation value is taken in the theory around the 
vacuum (\ref{irr decom}) in PWMM.
On the right-hand side of (\ref{result}), $M$ is a Hermitian 
matrix with the following block structure,
\begin{align}
M=\bigoplus_{s=1}^{\Lambda} ( M_s \otimes {\bf 1}_{N_5^{(s)}}),
\end{align}
where $M_s$ $(s=1,\cdots, \Lambda)$ are $N_2^{(s)}\times N_2^{(s)}$
Hermitian matrices.
$\langle \cdots \rangle_{MM}$ stands for an expectation value 
with respect to the following partition function,
\begin{align}
&Z_{{\cal R}}=\int \prod_{s=-\Lambda/2}^{\Lambda/2}
\prod_{i=1}^{N_2^{(s)}}dq_{si} Z_{\rm 1-loop}({\cal R},\{q_{si}\})
e^{-\frac{2}{g^2}\sum_{s}\sum_{i}N_5^{(s)}q_{si}^2},
\label{matrix model}
\end{align}
where ${\cal R}$ denotes the representation of (\ref{irr decom}),
$q_{si}$ are eigenvalues of $M_s$ and
\begin{align}
Z_{\rm 1-loop}=
\prod_{s,t=-\Lambda/2}^{\Lambda/2}
\prod_{J}
\prod_{i=1}^{N_2^{(s)}}\prod_{j=1}^{N_2^{(t)}}
\hspace{-5.5mm} {\phantom{\prod}}^{\prime}
\left[
\frac{\{(2J+2)^2+(q_{si}-q_{tj})^2\} \{(2J)^2+(q_{si}-q_{tj})^2\}}
{\{(2J+1)^2+(q_{si}-q_{tj})^2\}^2}
\right]^{\frac{1}{2}}.
\label{1loopdet}
\end{align}
In (\ref{1loopdet}), the product of $J$ runs from 
$|N_5^{(s)}-N_5^{(t)} |/2$ to $(N_5^{(s)}+N_5^{(t)})/2-1$.
And $\prod'$ means that the second factor in the numerator with 
$s=t$, $J=0$ and $i=j$ is not included in the product.

Note that (\ref{result}) implies that
the left hand side of (\ref{result}) does not depend on the positions 
$\{\tau_a \}$ of the operators.
This follows from the supersymmetry Ward identity in PWMM.
As shown in appendix A, there exists a fermionic matrix $\Psi_1$ 
in PWMM such that 
its variation under the supersymmetry is proportional to $D_1\phi$. 
Then, it follows from the Ward identity that
\begin{align}
0 = \delta_{s} \langle {\rm Tr} (\Psi_1 \phi^m )(\tau) \cdots \rangle
 = \frac{1}{m+1}\langle {\rm Tr} (D_1\phi^{m+1} )(\tau) \cdots \rangle
 = \frac{1}{m+1}
\frac{\partial}{\partial \tau}
\langle {\rm Tr}  (\phi^{m+1} )(\tau) \cdots \rangle,
\label{wi}
\end{align}
where $\cdots$ stands for operators made of $\phi$'s only.
Hence, they are indeed independent of the positions.
One can also check this easily by a perturbative calculation 
around each vacuum.

In this paper, we consider the case of $\Lambda=1$. 
In this case, (\ref{matrix model}) is just a one matrix model, 
\begin{align}
Z=
&\int \prod_{i}dq_{i}
\prod_{J=0}^{N_5-1}
\prod_{i>j}^{N_2}
\frac{\{(2J+2)^2+(q_i-q_j)^2\} \{(2J)^2+(q_i-q_j)^2\}}
{\{(2J+1)^2+(q_i-q_j)^2\}^2}
e^{-\frac{2N_5 }{g^2}\sum_{i} q_{i}^2}.
\label{lambda=1}
\end{align}
When $N_5=1$, (\ref{lambda=1}) takes the same form as 
the partition function of the 
six-dimensional version of the IKKT matrix model with a suitable regularization 
\cite{Kazakov:1998ji,Kitazawa:2006hq}.
%Unlike the 4 dimensional case studied in \cite{Kazakov:1998ji}, 
%it seems difficult to solve (\ref{lambda=1}) analytically. So we 
%apply a numerical method in the next section.
%
%If we take the D2 limit, (\ref{lambda=1}) is simplified to 
%\begin{align}
%Z_{D2}=
%&\int \prod_{i}^{N_2}dm_{i}
%\prod_{i>j}^{N_2} \tanh^2 \left( 
%\frac{\pi(m_i-m_j)}{2}
%\right)
%e^{-\frac{8\pi }{g^2}\sum_{i} m_{i}^2},
%\label{tanh}
%\end{align}
%where $g^2 = 4\pi g_{PW}^2 /N_5$ is the fixed parameter in the D2 limit.
%This model is identical with the partition function of the Chern-Simons 
%theory on $S^3$ with adjoint matter multiplets.
%One can solve this model in the strong coupling regime 
%in the planar limit \cite{Suyama:2011yz}. See appendix B.
%We discuss this model in Section~\ref{conclusion}.

%%%%%%%%%%%%%%%%%%%%%%%%%%%%%%%%%%%%%%%%%%%%%%%%%%%%%%%%%%%
\section{Gravity dual of PWMM}
\label{Gravity dual of PWMM}
%%%%%%%%%%%%%%%%%%%%%%%%%%%%%%%%%%%%%%%%%%%%%%%%%%%%%%%%%%%
In this section, we review the dual geometry for PWMM.
The gravity duals for the gauge theories with $SU(2|4)$ symmetry, 
which consist of PWMM, ${\cal N}=8$ SYM on $R\times S^2$ and 
${\cal N}=4$ SYM on $R\times S^3/Z_k$, were constructed
by Lin and Maldacena \cite{LM}\footnote{See also 
\cite{Ishiki:2006rt,Ishiki:2006yr}
for the gauge theory side.}.
They assumed the $SU(2|4)$ symmetric ansatz and then showed that 
finding the classical solutions is reduced to the problem of finding
axially symmetric solutions to the 3d Laplace equation
with appropriate boundary conditions given by parallel charged 
conducting disks and a background potential.

\subsection{Dual geometry of PWMM}
The supergravity solutions dual to the $SU(2|4)$ symmetric theories 
are given by 
\begin{align}
ds_{10}^2 &= 
\left( \frac{\ddot{V}-2\dot{V}}{-V''} \right)^{1/2}
\left\{
-4 \frac{\ddot{V}}{\ddot{V}-2\dot{V}}dt^2
-2 \frac{V''}{\dot{V}}(dr^2 +dz^2)
+4 d\Omega_{5}^2 +2 \frac{V'' \dot{V}}{\Delta} d\Omega_2^2
\right\}, \nonumber\\
C_1 &= - \frac{(\dot{V}^2)'}{\ddot{V}-2\dot{V}} dt, \;\;\;
C_3 = -4 \frac{\dot{V}^2 V''}{\Delta} dt \wedge d\Omega_2 , 
\nonumber\\
B_2 &= \left( 
\frac{(\dot{V}^2)'}{\Delta}+2z \right) d\Omega_2, \;\;\;
e^{4\Phi} = \frac{4(\ddot{V}-2\dot{V})^3}{-V'' \dot{V}^2 \Delta^2},
\label{LM solution}
\end{align}
where $\Delta = (\ddot{V}-2\dot{V})V''-(\dot{V}')^2$ and 
the dots and primes indicate $\frac{\partial}{\partial \log r} $
and $\frac{\partial }{\partial z}$, respectively.
Note that the solution is written in terms of a 
single function $V(r,z)$. 
The Killing spinor equations
in the supergravity are reduced to 
the Laplace equation for $V$ in a three-dimensional axially symmetric 
electrostatic system, where the coordinates for the axial and the 
transverse directions are given by $z$ and $r$, respectively.
Thus $V$ can be regarded as an electrostatic potential in this system. 
%In the following, we focus on the gravity dual of PWMM.

\begin{figure}
\begin{center}
\includegraphics[width=8cm]{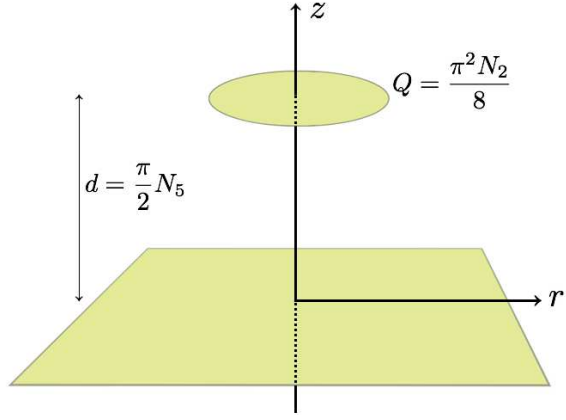}
\end{center}
\caption{The electrostatic system for PWMM around the vacua 
labeled by $(N_2,N_5)$.}
\label{figPWMM}
\end{figure}
The electrostatic system for the dual geometry of PWMM involves
an infinite conducting surface at $z=0$ and only 
the region $z\geq 0$ is relevant.
See Fig.~\ref{figPWMM}.  The positivity of the metric 
requires the presence of the background potential of the form 
$V_0( r^2z-\frac{2}{3}z^3)$, where $V_0$ is a constant.
In addition to the infinite surface, the system has some finite 
conducting disks. The positions and charges of these disks 
are related to the parameters of the vacua in PWMM.
For the gravity dual of PWMM around 
the vacuum (\ref{fuzzy sphere}) with (\ref{irr decom}), 
the system has $\Lambda$ disks each 
of which has the charge $\pi^2 N_2^{(s)}/8$ and resides
at the position $z=\pi N_5^{(s)}/2$, where $s=1,\cdots, \Lambda$.
(The radii of the disks are not free parameters. 
The regularity of the gravity solution demands that the 
charge density on a finite disk vanishes at the edge,
which relates the radius of the disk to the charge.)
The solution $V(r,z)$ of the Laplace equation in this electrostatic system 
determines the gravity solution (\ref{LM solution}) 
that is dual to the PWMM around the vacuum (\ref{fuzzy sphere}) 
with (\ref{irr decom}). 
In this geometry, $N_5^{(s)}$ and $N_2^{(s)}$ correspond to 
the charges of NS5-branes and D2-branes, respectively,
and $s=1,\cdots, \Lambda$ labels independent cycles 
with the flux of the branes.

The gravity solution that corresponds to the vacuum 
with $\Lambda=1$ in PWMM was studied in detail in \cite{Ling:2006up}.
The electrostatic system associated with this solution consists of one 
infinite conducting plate at $z=0$ and another finite 
conducting disk at $z=d>0$ with 
radius $R$ and charge $Q$. The background potential is given by
$V_0( r^2z-\frac{2}{3}z^3)$. 
$Q$ and $d$ are related to the brane charges as 
$N_5 =2d/\pi$ and $N_2=8Q/\pi^2$.
By solving the Laplace equation with these boundary conditions, 
one can determine the potential as 
\begin{align}
V_{PWMM}(r,z)&=V_0\left(r^2z-\frac{2}{3}z^3\right)
+V_0R^3\phi_{\kappa}(r/R,z/R), 
\label{VPWMM}
\end{align}
where $\kappa\equiv d/R$ and 
$\phi_\kappa(r,z)$ is given by
\begin{align}
\phi_\kappa(r,z)&=\frac{\beta(\kappa)}{\pi}\int^1_{-1}dt
\left(-\frac{1}{\sqrt{r^2+(z+\kappa+it)^2}}+\frac{1}{\sqrt{r^2+(z-\kappa+it)^2}}\right)
f_\kappa(t).
\label{pot}
\end{align}
Here $\beta(\kappa)$ is given in terms of 
$f^{(n)}_\kappa(t)$ defined in appendix A as
\begin{align}
\beta(\kappa)&\equiv 2\kappa\frac{f^{(2)}_\kappa(1)}{f^{(0)}_\kappa(1)},
\end{align}
and $f_{\kappa}(t)$ is the solution to the Fredholm integral equation 
of the second kind,
\begin{align}
f_\kappa(x)-\int_{-1}^1dyK_\kappa(x,y)f_\kappa(y)=1-\frac{f^{(0)}_\kappa(1)}{f^{(2)}_\kappa(1)}x^2
\label{f eq}
\end{align}
with kernel
\begin{align}
K_\kappa(x,y)=\frac{1}{\pi}\frac{2\kappa}{4\kappa^2+(x-y)^2}.
\label{K}
\end{align}
The equation (\ref{f eq}) is solved by 
\begin{align}
f_\kappa(t) = f^{(0)}_\kappa(t)-\frac{2\kappa}{\beta(\kappa)}f^{(2)}_\kappa(t)
=f^{(0)}_\kappa(t)
-\frac{f^{(0)}_\kappa(1)}{f^{(2)}_\kappa(1)}f^{(2)}_\kappa(t). 
\label{f}
\end{align}

%The constant potential on the disk ($r<R$) is 
%given as
%\begin{align}
%V_{PWMM}(r,d)\equiv V_0R^3\Delta(\kappa)
%=V_0R^3\left(\beta(\kappa)-\frac{2}{3}\kappa^3\right). \label{Delta}
%\end{align}
The charge density $\sigma_\kappa(r)$ for the radial direction 
on the disk is related to 
$f_\kappa(t)$ as
\begin{align}
\sigma_\kappa(r)=-\frac{\beta(\kappa)}{\pi^2}\int_r^1dt \frac{f'_\kappa(t)}{\sqrt{t^2-r^2}}, \;\;\;\;
f_\kappa(t)=\frac{2\pi}{\beta(\kappa)}\int_t^1 dr \frac{r \sigma_\kappa(r)}{\sqrt{r^2-t^2}}.
\end{align}
From this relation, 
one can interpret $f_{\kappa}(t)$ as the charge density projected onto
a diameter of the disk.
The radius of the disk is related to the charge as
\begin{align}
Q=q(\kappa)V_0 R^4, \;\;\;\;
q(\kappa)=\frac{\beta(\kappa)}{\pi}\int_{-1}^1 dt f_\kappa(t). \label{Q} 
\end{align}

The disk radius is related to the radius of $S^5$ at the 
edge of the disk as
\begin{align}
R= \frac{R^2_{S^5}}{4\alpha'}.
\label{R RS5}
\end{align}
One can easily check this by 
using the Laplace equation 
to rewrite $V''$ and note that $\dot{V}=0$ on the disk.

The parameters of the electrostatic problem were identified with 
the parameters in PWMM as \cite{LM, Ling:2006up}
\begin{align}
Q= \frac{\pi^2 N_2}{8},\hspace{5mm}  
d=\frac{\pi}{2}N_5, \hspace{5mm}  
R=\left(\frac{\pi^2 g^2 N_2}{m^3 hq(\kappa)}\right)^{\frac{1}{4}},\hspace{5mm}
V_0=\frac{h m^3 }{8g^2}.
\label{parameters}
\end{align}
Here, $h$ is a constant which does not depend on $g^2/m^3$, 
$N_2$ and $N_5$.
In section \ref{Emergent bubbling geometry}, 
we determine the value of $h$ from the 
gauge theory side.

\subsection{D2-brane limit}
\begin{figure}
\begin{center}
\includegraphics[width=8cm]{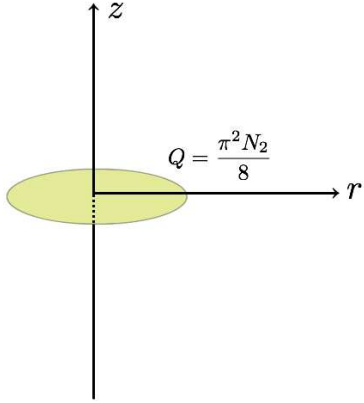}
\end{center}
\caption{The electrostatic system for the D2-brane solution.}
\label{figD2}
\end{figure}
The supergravity solution given by the potential (\ref{VPWMM}) has
two interesting scaling limits in which the 
solution becomes the D2-brane solution or
the NS5-brane solution constructed in \cite{LM}.
Let us first consider the limit to the D2-brane solution.
The D2-brane solution is given by the same form as (\ref{LM solution}).
The electrostatic system for this solution consists of 
a background potential 
\begin{align}
V_{D2} =W_0 (r^2 -2z^2)
\label{vd2}
\end{align}
with $W_0$ constant and a finite size disk at $z=0$ 
with charge $Q= \pi^2 N_2/8 $. 
Here, the system has no infinite surface and the
whole region of $z \in (-\infty, \infty)$ is considered as 
shown in Fig.~\ref{figD2}.
See \cite{LM} for the explicit form of this solution.

The D2-brane limit is given by redefining the coordinate $z \rightarrow d+z$
and focusing on the finite disk in the electrostatic system 
of (\ref{VPWMM}).
The limit is given as
\begin{align}
d\rightarrow \infty, \;\;\; Q: \;{\rm fixed},
\;\;\; V_0d =W_0 : \;{\rm fixed}.
\label{d2 limit gravity}
\end{align}
From (\ref{Q}) and (\ref{large k f}), we can see that 
this limit corresponds to the large-$\kappa$ limit.
After the redefinition $z \rightarrow d+z$, the background part of
(\ref{VPWMM}) becomes
\begin{align}
-\frac{2}{3}V_0 d^3 -2V_0d^2 z 
+V_0 d(r^2 -2z^2) +V_0 \left(zr^2 -\frac{2}{3}z^3 \right).
\label{expand V}
\end{align}
One can neglect the first and second terms since they do not affect
the supergravity solution which depends only on $\dot{V}, \ddot{V}, 
\dot{V}'$ and $V''$.
So up to these terms, (\ref{expand V}) indeed becomes 
(\ref{vd2}) in the limit of (\ref{d2 limit gravity}).

By using the relation (\ref{parameters}), one can 
rewrite this limit in terms of the parameters in PWMM as
\begin{align}
N_5 \rightarrow \infty, \;\;\; N_2: \;{\rm fixed}, \;\;\; 
\frac{4\pi g^2}{m^2N_5}\equiv g^2_{S^2} :\; {\rm fixed}.
\label{D2 limit gauge}
\end{align}
The limit corresponds to the commutative limit of fuzzy spheres, 
where PWMM describes $U(N_2)$ ${\cal N}=8$ SYM on $R\times S^2$.
The radius of $S^2$ is given by $1/m =1/2$.
The fixed quantity $g_{S^2}$ in (\ref{D2 limit gauge})
is the gauge coupling constant in this theory.

\subsection{NS5-brane limit}
\begin{figure}
\begin{center}
\includegraphics[width=8cm]{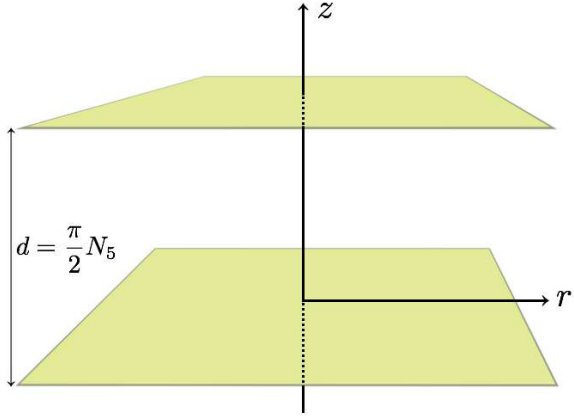}
\end{center}
\caption{The electrostatic system for the NS5-brane solution.}
\label{figNS5}
\end{figure}
Let us consider the NS5-brane limit, in which the gravity dual solution 
written in terms of (\ref{VPWMM}) is reduced to the NS5-brane solution 
constructed in \cite{LM}.
The NS5-brane solution is given by the form of (\ref{LM solution}),
where the electrostatic system now consists of 
two infinite conducting plates separated by 
distance $d$ as shown in Fig.~\ref{figNS5}.  
The electrostatic potential is given by
\begin{align}
V_{NS5}(r,z)=\frac{1}{g_0}\sin\left(\frac{\pi z}{d}\right) I_0 \left(\frac{\pi r}{d}\right), \label{VLST}
\end{align}
where $g_0$ is a constant and 
$I_n$ is the modified Bessel function of the first kind.
%The geometry given by this potential does not have a non-contractible 
%cycle for D2-branes and carries only NS5 brane charges.
%One can check that the solution has a single NS5-brane throat.
%The gravity solution that is associated with 
%NS5-branes on $S^5$. This solution 
%is considered as the gravity dual of type IIA LST on $R\times S^5$. 
For the explicit form of the geometry, see \cite{LM, Ling:2006up}.

The NS5-brane limit is given as a double scaling limit where
both $R$ and $V_0$ are sent to infinity in an appropriate way.
Let us review the derivation 
of the precise form of the scaling limit \cite{Ling:2006up}.
We first make the Fourier expansion of \eqref{VPWMM} in $r<R$ region as,
\begin{align}
  V_{PWMM}(r,z)=V_0R^3\Delta(\kappa)\frac{z}{d}
  +\sum_{n=1}^{\infty}c_n\sin\left(\frac{n\pi z}{d}\right)I_0\left(\frac{n\pi r}{d}\right),
  \label{VPWMM 2}
\end{align}
where $\kappa\equiv d/R$ and
$\Delta(\kappa)$ is defined as
\begin{align}
\Delta(\kappa)
=\beta(\kappa)-\frac{2}{3}\kappa^3.
\end{align} 
The restricted form of the expansion (\ref{VPWMM 2}) follows from 
the conditions that $V_{PWMM}$ is regular at $r=0$, 
constant ($V_0R^3\Delta(\kappa)$) at $z=d$ and zero at $z=0$.
Since the first term in (\ref{VPWMM 2}) does not contribute to the
geometry, the NS5-brane limit is a limit where
\begin{align}
c_1 \rightarrow \frac{1}{g_0}, \;\;\;\; c_n \rightarrow 0 \;\;  (n>1).
\label{c1 finite cn zero}
\end{align}
One can determine the coefficients $c_n$'s by 
the inverse Fourier transformation at $r=R$ as
\begin{align}
  c_n
%  &=\left(I_0\left(\frac{n\pi}{\kappa}\right)\right)^{-1}\frac{2}{d}V_0R^3
%  \int_0^ddz\left(\frac{z}{R}-\frac{2}{3}\left(\frac{z}{R}\right)^3
%+\phi_\kappa(1,z/R)
%  -\Delta(\kappa)\frac{z}{d}\right)\sin\frac{n\pi z}{d} \n
  =\left(I_0\left(\frac{n\pi}{\kappa}\right)\right)^{-1}2V_0R^3 p_n(\kappa),
\end{align}
where
\begin{align}
p_n(\kappa)=\int_0^1dy \left(\phi_\kappa(1,\kappa y)-\Delta(\kappa)y-\kappa y + \frac{2}{3}\kappa^3 y^3\right)\sin (n\pi y).
\end{align}
When $\kappa=d/R$ is small, $p_n(\kappa)$ behaves as 
\begin{align}
p_n(\kappa) \sim b_n \kappa^2,
\end{align}
where $b_n$ are constants. 
Since $I_n(z)\sim e^{z}/\sqrt{2\pi z}$ for $z\gg 1$,
we find for small $\kappa$ that
\begin{align}
c_n\sim 2b_n \sqrt{2\pi^2 n}e^{-\frac{n\pi R}{d}} V_0 (Rd)^{\frac{3}{2}}.
\end{align}
%and similarly $c_n$'s with $n>1$ have factors $e^{-n\pi /\kappa}$.
Then, the NS5-brane limit is given by
\begin{align}
R\rightarrow \infty, \hspace{5mm}  d:\; \text{fixed}, \hspace{5mm} 
V_0\rightarrow \frac{1}{g_0}\frac{1}{2b_1 \sqrt{2\pi^2}}(Rd)^{-\frac{3}{2}}e^{\frac{\pi R}{d}},
\label{NS5 limit gravity}
\end{align}
which realizes (\ref{c1 finite cn zero}).
Note that $\kappa=d/R$ goes to zero in this limit.
The value of $b_1$ was computed numerically 
and found to be $b_1=0.040$ \cite{Ling:2006up}.

Using the relations (\ref{parameters}), one can rewrite the limit 
\eqref{NS5 limit gravity} in the language of PWMM as
\begin{align}
N_2 \rightarrow \infty, \;\;\;
\lambda \rightarrow \infty, \;\;\; \frac{1}{N_2}
\lambda^{\frac{5}{8}}
e^{\frac{a}{N_5}\lambda^{\frac{1}{4}}} 
\equiv \tilde{g}_s :\; {\rm fixed}, \;\;\;
N_5 :\; {\rm fixed},
\label{NS5 limit gauge}
\end{align}
where $a=2\pi^{\frac{1}{2}}/h^{\frac{1}{4}}$ and $\lambda$ is the 
dimensionless 't Hooft coupling in PWMM, 
\begin{align}
\lambda = g^2N_2 \left(\frac{2}{m} \right)^{3}.
\end{align}
The dual theory of the NS5-brane solution is considered as 
a six-dimensional non-gravitational string theory called the 
little string theory (LST). The parameter $\tilde{g}_s$ is considered 
to be the string coupling constant of LST.
The limit (\ref{NS5 limit gauge}) predicts that 
the dynamics of PWMM near the NS5-brane limit
is controlled by $\lambda^{1/4}$.
We will confirm this in the next section by analyzing the gauge theory side.

%%%%%%%%%%%%%%%%%%%%%%%%%%%%%%%%%%%%%%%%%%%%%%%%%%%%%%%%%%%
\section{Emergent bubbling geometry}
\label{Emergent bubbling geometry}
%%%%%%%%%%%%%%%%%%%%%%%%%%%%%%%%%%%%%%%%%%%%%%%%%%%%%%%%%%%
In this section, we investigate the matrix integral 
(\ref{lambda=1}) in the parameter region where 
the dual supergravity description is valid.
In order for the supergravity approximation to be valid,
the brane charges, $N_2$ and $N_5$, should be very large 
and $N_2$ should be much larger than $\lambda$ and $N_5$
to suppresses the bulk string coupling.
In addition, it turns out that the condition 
$\lambda \gg N_5$ is needed to suppress the $\alpha'$ 
corrections.
We first show that the matrix integral \eqref{lambda=1} is
equivalent to a one-dimensional interacting Fermi gas model.
We then study the semi-classical limit of this model, which corresponds 
to the supergravity regime, 
by applying the Thomas-Fermi approximation.
Under this approximation, the system is described in terms of the 
mean-field density of the Fermi particles.
%which corresponds to 
%the eigenvalue density in the usual saddle point evaluation of 
%the matrix integral.
We find that the mean-field density can be identified with the 
charge density $f_\kappa$ in the gravity side.
We also solve the Fermi gas model in the D2-brane and NS5-brane limits 
and reproduce the radii of the geometries.

\subsection{Fermi gas model}
\label{Fermi gas model}
Here we show that the matrix integral (\ref{lambda=1}) can 
be mapped to a one-dimensional 
interacting Fermi gas system with $N_2$ particles.
We follow the method proposed in \cite{Marino:2012az}.

When $N_5$ is infinity, the measure factors in (\ref{lambda=1}) 
converge to $\tanh^2 \frac{\pi(q_i-q_j)}{2}$ up to
an over all constant.
The $1/N_5$ corrections are given as 
\begin{align}
 &\prod_{J=0}^{N_5-1}
 \frac{[(2J+2)^2+(q_i-q_j)^2][(2J)^2+(q_i-q_j)^2]}
 {[(2J+1)^2+(q_i-q_j)^2]^2}\nonumber \\
 &\quad = \tanh^2 \frac{\pi (q_i-q_j)}{2}
 \exp \left\{ \frac{2N_5}{(2N_5)^2+(q_i-q_j)^2}
 -\frac{2N_5[(2N_5)^2-3(q_i-q_j)^2]}{6[(2N_5)^2+(q_i-q_j)^2]^3}+\cdots 
\right\},
\label{N5 corrections}
\end{align}
where we have neglected the overall constant.
%We need lower-order terms of the expansion of $1/N_5$
%because the corresponding objects in the gravity side are 
%$N_5$ NS5-branes or D2-branes 
%and, in both cases, $N_5$ is considered as a large value.
%The Cauchy identity tells us that
By using the Cauchy identity, we rewrite the 
hyperbolic tangent part as
\begin{align}
 \prod_{i\neq j}^{N_2}
 \tanh \frac{\pi (q_i-q_j)}{2} 
 =\sum_{\sigma \in S_{N_2}}(-)^{\epsilon(\sigma)}
 \prod_{i=1}^{N_2}\frac{1}{\cosh \frac{\pi (q_i-q_{\sigma(i)})}{2}},
\end{align}
where $\epsilon(\sigma)$ stands for the sign of the permutation $\sigma$.

We introduce the operators, $\hat{p}$ and $ \hat{q}$, that
obey the canonical Heisenberg algebra $[\hat{p}, \hat{q}]=-i$.
Let ${\cal H}$ be the usual representation space of this algebra 
which is an infinite dimensional Hilbert space spanned by the 
eigenstates of $\hat{q}$. We denote the eigenstates by 
$| q\rangle$, which satisfy $\hat{q}|q \rangle = q |q \rangle$.
Then, we have
\begin{align}
 \frac{1}{\cosh \frac{\pi (q_i-q_{j})}{2}}
=\frac{1}{\pi}\int dp\,\frac{1}{\cosh p}e^{ip(q_i-q_j)}
=\left\langle q_i\left| \frac{2}{\cosh \hat{p}}\right| q_j\right\rangle .
\end{align}
We also introduce the Hilbert space for $N_2$ Fermions. 
It is a subspace of ${\cal H}^{\otimes N_2}$
and spanned by the antisymmetric states, 
\begin{align}
 \left| q_1,\cdots ,q_{N_2}\right\} :=
 \frac{1}{N_2!}\sum_{\sigma \in S_{N_2}}(-)^{\epsilon(\sigma)}
\left| q_{\sigma(1)} \right\rangle
\otimes
\left| q_{\sigma(2)} \right\rangle
\otimes \cdots
\otimes
\left| q_{\sigma(N_2)}\right\rangle.
\label{Fermi states}
\end{align}
We denote by $\hat p_i$ and $\hat q_i$ the 
canonical pair 
on the $i$-th Hilbert space.
They obey the commutation relations, 
$[\hat p_i,\hat q_j]=-i\delta_{ij}$.
With these notations, we can rewrite 
the matrix integral (\ref{lambda=1}) 
as the partition function of 
a Fermi gas system,
\begin{align}
 Z&= \Tr \hat \rho 
\label{Fermipf}
\end{align}
where 
the trace is taken over the states (\ref{Fermi states}) as
\begin{align}
\Tr \hat \rho =
\int \prod_{i}dq_{i}
 \left\{ q_1,\cdots ,q_{N_2}\left| \hat \rho \right| q_1,\cdots ,q_{N_2}
\right\},
\end{align}
and the density matrix is given by
\begin{align}
 \hat \rho=
 \prod_{i=1}^{N_2}e^{-U(\hat q_i)/2}
 \prod_{i\neq j}^{N_2}e^{-\frac{1}{4}W(\hat q_i-\hat q_j)}
 \prod_{i=1}^{N_2}e^{-T(\hat p_i)}
 \prod_{i=1}^{N_2}e^{-U(\hat q_i)/2}
 \prod_{i\neq j}^{N_2}e^{-\frac{1}{4}W(\hat q_i-\hat q_j)}.
\label{density}
\end{align}
%with a trivial energy shift %, $N\log (2(N-1)!)$, 
%omitted.
The functions $T(x)$, $U(x)$ and $W(x)$ are defined as follows.
\begin{align}
 &T(x):=\log \cosh x,\nonumber \\
 &U(x):=\frac{2N_5}{g^2}x^2,\nonumber \\
 &W(x):=-\frac{2N_5}{(2N_5)^2+x^2}.
\end{align}
Here, we have kept only the first term of
the exponent in (\ref{N5 corrections}) 
because we are interested in the large-$N_5$ limit.
Even if $N_5$ is large, the first term should be kept since 
it can become comparable to the Gaussian potential in 
some parameter regions.
The model defined by (\ref{Fermipf})
is an interacting one-dimensional Fermi gas system
of $N_2$ fermions, where 
the interaction is given by $W(q_i-q_j)$.

The semi-classical limit of this model is 
described by the many-body Hamiltonian, 
\begin{align}
\hat{H} = \sum_{i} T(\hat{p}_i)  +\sum_{i} U(\hat{q}_i)
+\frac{1}{2}\sum_{i\neq j} W(\hat{q}_i- \hat{q}_j).
\label{scH}
\end{align}
When $N_2$ is large, we can apply 
the Thomas-Fermi approximation at zero temperature 
(see in appendix C) to the system (\ref{scH}).
In this approximation, the original many-body path integral 
can be evaluated at a saddle point characterized by the
mean-field density $\rho(q)$.
$\rho(q)$ is assumed to have a single support $[-q_m, q_m]$ and 
it is normalized as
\begin{align}
 \int_{-q_m}^{q_m}dq\, \rho(q)=N_2. 
\label{rho normalization}
\end{align}
$\rho(q)$ is determined by (\ref{integral equation}) which follows 
from the Thomas-Fermi equation at zero temperature.
In our case, the equation (\ref{integral equation}) is given by
\begin{align}
\mu =\pi \rho(q)+\frac{2N_5}{g^2}q^2
 -\int_{-q_m}^{q_m}dq'\, \frac{2N_5}{(2N_5)^2+(q-q')^2}\rho(q'),
\label{eq:atFermiSurface}
\end{align}
where $\mu$ is the chemical potential.
Here, we have made an approximation that $T(p)=\log \cosh p \sim |p|$.
This is valid when $N_2$ is large.

The equation (\ref{eq:atFermiSurface}) can also be obtained 
from the usual saddle-point analysis for matrix integrals,
where $\rho(q)$ is interpreted as the eigenvalue density.
By noting that $\log \tanh^2 \frac{\pi q_m x}{2} \rightarrow - \pi \delta(x) $
as $q_m \rightarrow \infty$, one can see that
(\ref{eq:atFermiSurface}) is just a saddle-point equation for the 
eigenvalue density and $\mu$ plays the role of the Lagrange multiplier
which imposes the normalization (\ref{rho normalization}). 
So the semi-classical equation (\ref{eq:atFermiSurface}) is 
expected to be valid when $q_m \gg 1$ in the large-$N_2$ limit. 
We will see in section \ref{Range of the semi-classical limit} that
the quantum corrections in the Fermi gas model are 
indeed negligible when $q_m \gg 1$.
We will also see that the condition $q_m \gg 1$
is written as $\lambda \gg N_5$ in terms of the original
parameters in PWMM. 
This is a strong coupling region of PWMM and 
corresponds to the region in 
the gravity side where the $\alpha'$ corrections are negligible.

%At weak coupling as $\lambda \ll N_5$, 
%the Gaussian atractive force becomes strong 
%and the system is equivalent to a simple Gaussian matrix integral.
%Here, the mean-field density is localized around the origin because of the 
%attractive force, so that the condition $q_m \gg 1$ is not satisfied. 
%At strong coupling, however, the Gaussian force becomes weaker and  
%the support of the mean-field density becomes larger, namely, $q_m \gg 1$.
%Then the system can be treated semi-calassically and
%it is described by the integral equation (\ref{eq:atFermiSurface}). 
%This is exactly the region where the 
%Fermi gas system describes the electrostatic problem 
%on the gravity side as we will see below.

%%%%%%%%%%%%%%%%%%%%%%%%%%%%%%%%%%%%%%%%%
\subsection{Mapping to the gravity side}
%%%%%%%%%%%%%%%%%%%%%%%%%%%%%%%%%%%%%%%%%
The integral equation \eqref{eq:atFermiSurface}
for the mean-field density $\rho(q)$  
in the semi-classical limit 
is the same type as the equation \eqref{f eq} for the charge density 
$f_{\kappa}(x)$ in the gravity dual.
So we propose the following identification.
\begin{align}
\rho(q)=\frac{\mu}{\pi}f_{\kappa}(q/q_m), \;\;\;
\frac{N_5}{q_m}=\kappa.
\label{rho f}
\end{align}
Under this identification, \eqref{eq:atFermiSurface} 
is completely equivalent to \eqref{f eq}.
In the following, we make consistency checks of the
relation (\ref{rho f}).

Based on the relation (\ref{rho f}),
we first translate the parameters
in PWMM to those in the gravity side as follows.
First, since $\kappa$ is related to the radius of the disk
as $\kappa=d/R=\pi N_5/2R$, we have
\begin{align}
q_m=\frac{2}{\pi}R. 
\label{qm R}
\end{align}
From the integral equation, we also have
\begin{align}
\frac{2N_5q_m^2}{\mu g^2}=\frac{f^{(0)}_{\kappa}(1)}{f^{(2)}_{\kappa}(1)}.
\label{qm}
\end{align}
Then, by comparing (\ref{Q}) and \eqref{rho normalization}, 
we obtain
\begin{align}
\frac{N_2}{\mu q_m}=\frac{q(\kappa)}{\beta(\kappa)}. 
\label{mu}
\end{align}
Finally, from (\ref{qm}) and (\ref{mu}), we obtain
\begin{align}
\frac{\lambda}{q_m^4}&=q(\kappa), \label{qm2} \\
\frac{\mu \lambda}{N_2q_m^3}&=\beta(\kappa), \label{mu2}
\end{align}
where $\lambda=g^2N_2$ is the 't Hooft coupling of PWMM.
From equations (\ref{rho f}) and \eqref{qm2}, 
we find that $\kappa$ depends only on the combination of $\lambda/N_5^4$ 
and $q_m$ depends only on $\lambda$ and $N_5$.

It should be noted that the relation \eqref{qm R} is consistent 
with the fact that $h$ in \eqref{parameters} is a constant and independent 
of $N_2$, $N_5$ and $\lambda$. 
In fact, from \eqref{Q} and \eqref{parameters} with \eqref{qm R}, 
one can determine $h$ as $h=2/\pi^2$.
Thus the constant $a=2\pi^{\frac{1}{2}}/h^{\frac{1}{4}}$ 
in \eqref{NS5 limit gauge} is given by
\begin{align}
a=2^{\frac{3}{4}}\pi.
\end{align}

In the following, we consider the D2-brane limit \eqref{D2 limit gauge}
and the NS5-brane limit \eqref{NS5 limit gauge},
which correspond to the large-$\kappa$ limit and 
the small-$\kappa$ limit, respectively.
When $\kappa$ is large, the term with the integral kernel in 
\eqref{eq:atFermiSurface} is negligible and 
the system becomes just a set of $N_2$ free fermions.
On the other hand, when $\kappa$ is small, 
the effective interactions between the fermions become very strong\footnote{
This can be seen as follows.
$q_m$ can be considered as a typical length scale of the system 
and then the effective interaction potential is given by
$\tilde{W}(y):=q_m W(q_my)=-\frac{\kappa}{\kappa^2+y^2}$.
Hence the interaction range and the force are proportional to
$\kappa$ and $1/\kappa^2$, respectively. 
%So the small-$\kappa$ region corresponds to
%a strong interaction with a short range.
}.
In these two limits, 
\eqref{eq:atFermiSurface} is solvable
and we can find solutions for $\rho(x)$ and $q_m$. 
Then from the relations (\ref{R RS5}) and (\ref{qm R}), 
we can compute the radii of $S^5$'s as the
range of the mean-field density as\footnote{Here we put $\alpha'=1$.},
\begin{align}
R_{S^5}^2 = 2\pi q_m.
\label{RS5andqm}
\end{align}
We will show that the radii obtained from the Fermi gas model
through (\ref{RS5andqm}) agree with known results 
obtained from the gravity solutions \cite{LM}.

\subsubsection*{D2-brane limit (large-$\kappa$ limit)}
On the gravity side, 
D2-brane limit corresponds to the limit of large-$\kappa$.
In this limit, we can find solutions for
$\beta(\kappa)$, $q(\kappa)$ and $f_\kappa(q)$ by solving the 
integral equation (\ref{f eq}).
The solutions are given by \eqref{large k f} in appendix A.

From the relations, \eqref{rho f}, \eqref{qm2} and \eqref{mu2},
one can then obtain solutions for $\rho(q)$, $q_m$ and $\mu$ 
in the Fermi gas model as
\begin{align}
\rho(q)&=N_2\left(\frac{9N_5}{8\pi\lambda}\right)^{\frac{1}{3}}\left[1-\left(\frac{q}{q_m}\right)^2\right], \;\;\;
q_m=\left(\frac{3\pi \lambda}{8N_5}\right)^{\frac{1}{3}}, \;\;\;
\mu=N_2\left(\frac{9\pi^2N_5}{8\lambda}\right)^{\frac{1}{3}}.
\label{solution in D2}
\end{align}
The parameter $\kappa$ on the gravity side is given by 
\begin{align}
\kappa&=\frac{2}{(3\pi)^{\frac{1}{3}}}
\left(\frac{N_5^4}{\lambda}\right)^{\frac{1}{3}}.
\label{kappa D2} 
\end{align}
So the D2-brane limit (\ref{D2 limit gauge}) in PWMM  
is indeed mapped to the limit of $\kappa \rightarrow \infty$
through the identification (\ref{rho f}).

In the D2-brane limit, PWMM is reduced to 
$U(N_2)$ $\mathcal N=8$ SYM on $R\times S^2$. 
The gravity dual of $\mathcal N=8$ SYM on $R\times S^2$ around the 
trivial vacuum is constructed in \cite{LM}.
It is shown that the radius of $S^5$ at the edge of the disk
is related to the 't Hooft coupling in the SYM as
\begin{align}
R_{S^5}^2=\pi \left(3 g^2_{S^2}N_2 \right)^{\frac{1}{3}}. 
\label{S5 radius D2}
\end{align}
Here, the coupling constant $g_{S^2}$ in SYM is
related to that in PWMM as shown in (\ref{D2 limit gauge}).
On the other hand, we can compute the $S^5$ radius in the D2-brane limit
from our solutions (\ref{solution in D2}). 
By substituting the solution (\ref{solution in D2}) for $q_m$ 
to the relation (\ref{RS5andqm}), 
we can reproduce (\ref{S5 radius D2}).
This shows the consistency of our identification (\ref{rho f}).

By using the solution (\ref{solution in D2}) of the Fermi gas model,
one can compute correlators in this limit. 
For example, the vev of the loop operator is given by
\begin{align}
%\left\langle\frac{1}{N_2N_5}\mathrm{Tr}e^{-il\phi/2}\right\rangle
%&=\frac{1}{N_5}\mathrm{Tr}_{N_5\times N_5}e^{-2l i L_4^{[N_5]}}
%\times\frac{1}{N_2}\int_{-q_m}^{q_m}dq \rho(q)e^{lq} \n
%&=\frac{1}{N_5}\frac{\sin lN_5}{\sin l}\cdot \frac{3}{l^3q_m^3}\left\{(lq_m)\cosh(lq_m)-\sinh(lq_m)\right\}.
\frac{1}{N_2N_5}\left\langle\mathrm{Tr}e^{lM}\right\rangle_{MM}
=\frac{1}{N_2}\int_{-q_m}^{q_m}dq \rho(q)e^{lq}
=\frac{3}{l^3q_m^3}\left\{(lq_m)\cosh(lq_m)-\sinh(lq_m)\right\}.
\label{loop D2}
\end{align}
The free energy $F$ defined in \eqref{F2} is given by
\begin{align}
%F=-\frac{1}{2}\left(\frac{9\pi^2 N_5}{8\lambda}\right)^{\frac{1}{3}}N_2^2.
F=-\frac{9}{10}\left(\frac{\pi^2N_5}{3\lambda}\right)^{\frac{1}{3}}N_2^2.
\end{align}
Note that the free energy is a generating function of 
$\mathrm{Tr}M^2$. 
Correlation functions of this operator can be 
computed as derivatives of the free energy. For example, 
\begin{align}
\frac{1}{N_2N_5}\langle \mathrm{Tr}M^2\rangle_{MM} =-\frac{1}{2N_2^2N_5}\frac{\partial}{\partial (1/\lambda)}F
%=\frac{(9\pi^2)^{\frac{1}{3}}}{24}\left(\frac{\lambda}{N_5}\right)^{\frac{2}{3}},
=\frac{(9\pi^2)^{\frac{1}{3}}}{20}
\left(\frac{\lambda}{N_5}\right)^{\frac{2}{3}}.
\end{align}
Of course, this agrees with the quadratic term in $l$ in (\ref{loop D2}).
%Multi-point functions can also be computed as higher
%derivatives of the free energy.

One can also solve the path integral (\ref{lambda=1}) in this limit 
by applying the usual saddle-point technique.
See appendix~\ref{Solving the D2 brane limit at the planar level}.

\subsubsection*{NS5-brane limit (small-$\kappa$ limit)}

On the gravity side, the NS5-brane limit 
is a limit of small $\kappa$. When $\kappa$ is small, 
$\beta(\kappa)$, $q(\kappa)$ and $f_\kappa(q)$ are 
solved by \eqref{small k f} in appendix A.

In this limit, from \eqref{rho f}, \eqref{qm2} and \eqref{mu2}, 
we obtain solutions $\rho(q)$, $q_m$ and $\mu$ for the 
Fermi gas model as
\begin{align}
\rho(q)=\frac{8^{\frac{3}{4}}N_2}{3\pi\lambda^{1/4}}\left[1-\left(\frac{q}{q_m}\right)^2\right]^{\frac{3}{2}}, \;\;
q_m=(8\lambda)^{\frac{1}{4}}, \;\;
\mu=\frac{8^{\frac{1}{2}}N_2N_5}{\lambda^{\frac{1}{2}}}.
\label{solution NS5}
\end{align}
The parameter $\kappa$ on the gravity side 
is given by 
\begin{align}
\kappa&=\frac{N_5}{(8\lambda)^{\frac{1}{4}}}.
\end{align}
So the NS5-brane limit (\ref{NS5 limit gauge}) in PWMM, 
which implies $\lambda\sim N_5^4 (\log N_2)^4 \gg N_5^4$, 
is consistently mapped to 
the small-$\kappa$ limit. Furthermore, 
we can see from (\ref{solution NS5})
that the typical scale of the mean-field density is 
given by $\lambda^{1/4}$. 
So the dynamics in this regime is governed by $\lambda^{1/4}$ as 
mentioned in the last section.

The gravity dual of PWMM in the small-$\kappa$ limit is studied in \cite{LM}.
The radius of $S^5$ at the edge of the disk is given as\footnote{
Note that $m$ in the equation (D8) in \cite{LM} 
is related to our mass parameter $m$ as $m_{LM}=m_{ours}/2$.}
\begin{align}
R_{S^5}^2=2\pi(8\lambda)^{\frac{1}{4}}. \label{S5 radius PWMM}
\end{align}
Again we can reproduce this result from the solution 
(\ref{solution NS5}) in the Fermi gas model, by using 
(\ref{RS5andqm}). 
This gives another supporting evidence for 
our identification (\ref{rho f}).

The vev of the loop operator in this limit is given by
\begin{align}
%\left\langle\frac{1}{N_2N_5}\mathrm{Tr}e^{-il\phi/2}\right\rangle
%=\frac{1}{N_5}\frac{\sin lN_5}{\sin l}\cdot 
%\frac{2\sqrt{2}}{l^2}\lambda^{-\frac{1}{2}}I_2\left[l(8\lambda)^{\frac{1}{4}}\right].
\frac{1}{N_2N_5}\left\langle\mathrm{Tr}e^{lM}\right\rangle_{MM}
=\frac{2\sqrt{2}}{l^2}\lambda^{-\frac{1}{2}}I_2\left[l(8\lambda)^{\frac{1}{4}}\right].
\end{align}
The free energy $F$ is given by
\begin{align}
%F=-\frac{\sqrt{2}N_5}{\lambda^{\frac{1}{2}}}N_2^2.
F=-\frac{4\sqrt{2}N_5}{3\lambda^{\frac{1}{2}}}N_2^2.
\end{align} 
The vev of $\mathrm{Tr}M^2$ can also be computed as
\begin{align}
\frac{1}{N_2N_5}\langle \mathrm{Tr}M^2\rangle_{MM} =-\frac{1}{2N_2^2N_5}\frac{\partial}{\partial(1/\lambda)}F
%=\frac{\sqrt{2}}{4}\lambda^{\frac{1}{2}}
=\frac{\sqrt{2}}{3}\lambda^{\frac{1}{2}}.
\end{align}

%%%%%%%%%%%%%%%%%%%%%%%%%%%%%%%%%%%%%%%%%%%%%%%%%%%%%%%%%%%%%%%%
\subsection{Range of the semi-classical approximation}
\label{Range of the semi-classical limit}
%%%%%%%%%%%%%%%%%%%%%%%%%%%%%%%%%%%%%%%%%%%%%%%%%%%%%%%%%%%%%%%
In this subsection, we consider when the semi-classical 
approximation used in section \ref{Fermi gas model} is valid.
We first expand the partition function (\ref{Fermipf}) around 
the semi-classical limit by using the Wigner transformation.
The Wigner transform of an operator $\hat A$ on ${\cal H}$ 
is defined by
\begin{align}
 A_{W}(q,p)=\int dq'\, \left\langle q-\frac{q'}{2}\left| \hat A \right| q+\frac{q'}{2}\right\rangle e^{ipq'}.
\end{align}
It is easy to see that the 
product of two operators $\hat A$ and $\hat B$ is translated 
to the $\star$-product of their Wigner transforms: 
\begin{align}
 (\hat A\hat B)_{W}=A_{W}\star B_{W},
\end{align}
where the $\star$-product is defined by
\begin{align}
(f \star g)(p,q)
=
f(p,q)
\exp \left[ \frac{i}{2}(\overleftarrow{\partial}_q\overrightarrow{\partial}_p-\overleftarrow{\partial}_p\overrightarrow{\partial}_q)\right]
g(p,q).
\label{eq:starprod}
\end{align}
The trace is written as the integral over the phase space:
\begin{align}
 \Tr \hat A=\int \frac{dqdp}{2\pi}A_W(q,p).
\end{align}
Hence, the partition function (\ref{Fermipf}) becomes
\begin{align}
 Z=\Tr \hat\rho
 =\int \prod_i \frac{dq_idp_i}{2\pi}\rho_W .
\end{align}
We define
\begin{align}
 &\mathcal{U}_i:=U(q_i)+\frac{1}{2}
\sum_{j\neq i}^{N_2}W(q_i-q_j).
\end{align}
Then using the Baker-Campbell-Hausdorff formula, 
we obtain the expansion around the semi-classical limit as
\begin{align}
 \rho_W
 &=e^{-\sum_{i}\mathcal{U}_i/2}\star e^{-\sum_{i}T(p_i)}\star e^{-\sum_{i}\mathcal{U}_i/2}\nonumber \\
 &=\exp \left\{\textstyle
 -\sum_i\left( T(p_i)+\mathcal{U}_i\right) 
 -\frac{1}{24}\left[\sum_i \left(2T(p_i)+\mathcal{U}_i\right),
\left[ \sum_{j}T(p_j),\sum_{k}\mathcal{U}_k\right]_\star\right]_\star+\cdots
%\nonumber \\
% &\quad\qquad\qquad
% +\frac{1}{12}\left[ \sum_{i}\Big( T(p_i)-\mathcal{U}_i\Big) ,
% \left[\sum_{j}T(p_j),\sum_{k}\mathcal{U}_k\right]_\star \right]_\star
% +\cdots
% \Big) 
\right\} ,
\label{eq:BCH}
\end{align}
where the star commutator is defined by 
$[f, g]_{\star}=f\star g - g\star f$.
Note that the density matrix is Hermitian, 
so that only terms with the even number of commutators appear in the right-hand side.

Then let us consider when we can neglect the correction terms 
in (\ref{eq:BCH}) that have the star commutators.
We always assume the large-$N_2$ limit where 
the saddle-point configurations dominate.
Let us denote the orders (magnitudes) of 
$q_i$, $p_i$, $T(q_i)$ and $\mathcal{U}_i$
at the saddle point by $q$, $p$, $T$ and 
$\mathcal{U}$, respectively. 
(From the symmetry, the order should be the same 
for any $i=1,\dots,N_2$.)
These orders are related as follows.
First, at the saddle point, the kinetic energy and the 
potential energy should be in equilibrium. 
So we have $T={\mathcal U}$.
Secondly, in the large-$N_2$ limit, 
$T(p_i)$ is approximated by $|p_i|$, so that we also have $T=p$.
Thirdly, since the Fermi momentum 
is related to $\rho(q)$ by (\ref{rho and pf}) 
and $\rho(q)$ satisfies the normalization 
(\ref{rho normalization}), it follows that $p=N_2/q$.
Thus we have
\begin{align}
T = {\mathcal U} = p = N_2/q.
\end{align}
Note that $pq=N_2$ is large.
The semi-classical part (the first term of the exponent in 
(\ref{eq:BCH})) looks like the order of $N_2(T+{\mathcal U}) \sim N_2^2/q$. 
However, this is not true in general. Near the NS5-brane limit, 
there is a cancellation between $\sum_iT(p_i)$ and 
$\frac{1}{2}\sum_{i\neq j}W(q_i-q_j)$, 
so the order is given by $\sum_i U(q_i)\sim N_2^2N_5q^2/\lambda$.
The order of the commutator term,
$\left[\sum_iT(p_i),\left[ \sum_jT(p_j),\sum_k\mathcal{U}_k \right]_{\star}\right]_\star$,
is given by $N_2T^2\mathcal{U}/(pq)^2 \sim N_2^2/q^3$.
Hence, if
\begin{align}
q^5 \gg \frac{\lambda}{N_5}, 
\label{cond-sca2}
\end{align}
the commutator term is negligible.
Then let us consider the higher order terms.
Note that, since $T$ is approximated by $p$ in our limit, 
the commutator terms with more than one $\mathcal{U}$
in \eqref{eq:BCH} vanish.
Then, the general higher order terms take the form of
$[T,[T,[\cdots,[T,\mathcal{U}]\cdots]]]$.
The order of such a term with $2m$ $T$'s and one $\mathcal{U}$
is estimated as $N_2\mathcal{U}/q^{2m}\sim N_2^2 /q^{2m+1}$. 
So these terms are suppressed compared to the term with a single 
commutator, if 
\begin{align}
q \gg 1 .
\label{qgg1}
\end{align}
So we find that when (\ref{cond-sca2}) and (\ref{qgg1}) are 
satisfied, the semi-classical approximation is valid.

However, it turns out that one of the two conditions,
(\ref{cond-sca2}) and (\ref{qgg1}), is redundant, namely, 
they are equivalent to each other for the saddle-point configurations.
In fact, we can see that both of these conditions
are equivalent to the following single condition which is written 
in terms of the original parameters in PWMM.
\begin{align}
\lambda \gg N_5.
\label{cond-sca3}
\end{align}
For example, one can show that (\ref{cond-sca3}) follows 
from (\ref{cond-sca2}) as follows.
In the D2-brane limit, $q_m$ is given in (\ref{solution in D2}).
So (\ref{cond-sca2}) implies (\ref{cond-sca3}). 
Similarly, the NS5-brane limit corresponds to 
$ 1 \gg \kappa =N_5/q_m \sim N_5/\lambda^{1/4} $ which 
implies (\ref{cond-sca3}). 
(Note that we always assume $N_5 \gg 1$.)
When $\kappa$ is in the intermediate region, 
(\ref{cond-sca3}) should also be satisfied, 
because $q(\kappa)$ is a smooth positive function of $\kappa$.
In fact, if $\kappa =\frac{N_5}{q_m} \sim 1$, we have 
$\lambda \sim N^4_5$ from (\ref{qm2}). This implies (\ref{cond-sca3}). 
In the same way, one can derive (\ref{cond-sca2}) from 
(\ref{qgg1}), and (\ref{qgg1}) from (\ref{cond-sca3}).
Therefore, the three conditions, (\ref{cond-sca2}), (\ref{qgg1}) and
(\ref{cond-sca3}), are all equivalent.

From (\ref{RS5andqm}), we see that
the condition (\ref{qgg1}) corresponds to the case of 
large radius of the $S^5$.
So the semi-classical regime of the Fermi gas system
corresponds consistently to 
the region in the gravity side in which we can neglect 
the $\alpha'$ corrections.

%%%%%%%%%%%%%%%%%%%%%%%%%%%%%%%%%%%%%%%%%%%%%%%%%%%%%%%%%%%%%
\section{Summary and discussion}
\label{Summary and discussion}
%%%%%%%%%%%%%%%%%%%%%%%%%%%%%%%%%%%%%%%%%%%%%%%%%%%%%%%%%%%%%
The gravity dual of the plane wave matrix model (PWMM) is 
given by bubbling geometries in type IIA superstring theory. 
These geometries are associated with the problem of 
an axially symmetric electrostatic system with some conducting disks 
and an appropriate background potential, which is defined on 
a certain two-dimensional subspace of the ten-dimensional space-time.
The solution is written only in terms of a single function, 
which corresponds to
the electrostatic potential in the axially symmetric system.
Once one finds the potential by solving the Laplace equation of the system,
one can construct the corresponding solution in the 
ten-dimensional supergravity.
In this paper, we studied an emergent phenomena for this geometry by 
investigating a quarter BPS sector of the plane wave matrix model 
that is associated with the field $\phi$ defined in (\ref{phi}). 
Since $\phi$ is a complex field and 
has two real degrees of freedom, the emergent geometry described 
in this sector is expected to be a two-dimensional surface ${\cal M}_{\phi}$.
We identified ${\cal M}_{\phi}$ with the two-dimensional surface 
on the gravity side on which the electrostatic problem is defined.
%When the time coordinate is Wick-rotated to the Lorentzian signature, 
%the field $\phi$ breaks the original $R\times SO(3)\times SO(6)$
%to $R\times SO(2)\times SO(5)$. 
%Then from the symmetry, ${\cal M}_{\phi}$ should be fibered on 
%$R\times SO(3)\times SO(6)/ (SO(2)\times SO(5))$.
%(In other words, the definition of 
%$\phi$ picks up a point on $R\times S^2 \times S^5$ in 
%the dual gravity picutre and ${\cal M}_{\phi}$ is defined on that point.
%From the original symmetry, ${\cal M}_{\phi}$ should exists everywhere
%on $R\times S^2 \times S^5$.)  So the topology of the total space 
%is locally given by $R\times S^2 \times S^5 \times {\cal M}_{\phi}$.
%On the other hand, the topology of the gravity dual solution of PWMM is
%$R\times S^2 \times S^5 \times {\cal M}_e$, where ${\cal M}_e \sim R^2$ 
%is a two dimensional plane. ${\cal M}_e$ is the space on which 
%the axially symmetric electrostatic system is defined.
%By comparing these topological structures, we made an identification that 
%${\cal M}_{\phi}$ corresponds to ${\cal M}_e$, namely, 
%the emergent space for the sector of $\phi$ is the space for 
%the electrostatic system in the gravity dual solution. 

We considered PWMM around the vacua (\ref{fuzzy sphere}) 
with (\ref{irr decom}).
We applied the localization method to the sector of $\phi$
and obtained a matrix integral. 
We investigated the case with $\Lambda=1$ and
mapped the matrix integral to a one-dimensional 
interacting Fermi gas system. 
And then we applied the Thomas-Fermi approximation which is 
valid in the semi-classical limit.
We found that the mean-field density of the Fermi particles 
satisfies the same integral equation as the 
disk charge density in the electrostatic 
problem on the gravity side.
Then, we proposed the identification (\ref{rho f}) of these two objects.
Since the whole geometry can be reconstructed from the 
charge density, this relation gives a realization of the emergent geometry
in PWMM.
%The geometry of the mean-field density is trivial 
%at weak coupling since the path integral reduces to 
%a matrix Gaussian integral. However, it describes 
%the dual geometry described by the charge density  
%emerges in the strong coupling limit

We made some consistency checks of our identification and 
obtained positive results. 
We consider two scaling limits, the D2-brane limit and 
the NS5-brane limit. In these limits, the gravity dual of PWMM is 
reduced to the solutions associated with the corresponding branes.
We found that the D2-brane and the NS5-brane limits correspond to the
free and the strongly coupled limits in the Fermi gas system, respectively.
By solving the Fermi gas system in these limits, we found 
the value of $a$ in (\ref{NS5 limit gauge}), which 
was not fixed solely by the argument in \cite{Ling:2006up}.
We reproduced the radius of $S^5$'s in the D2-brane and the NS5-brane
geometries in terms of the solutions of the Fermi gas model.
These results strongly support our identification between 
the mean-field density and the charge density.
%Our result also shows that the dynamics of PWMM in the NS5-brane limit
%is controlled by the effective coupling $\lambda^{1/4}/N_5$.
%This is consistent with the gravity side (\ref{NS5 limit gauge}).
In particular, our result in the NS5-brane limit reproduces 
the known behavior of the fivebrane radius 
proportional to $\lambda^{1/4}$\cite{Maldacena:2002rb}
and it gives a strong evidence for the description of 
fivebranes in PWMM proposed in \cite{Maldacena:2002rb}.

There remain some problems that are not considered in this paper.
First, our analysis in the NS5-brane limit is valid only in the planar 
limit. It corresponds to the leading order of the string coupling constant,
$\tilde{g}_s$ defined in (\ref{NS5 limit gauge}), in the little string theory.
However, the existence of the NS5-brane limit (\ref{NS5 limit gauge}) in PWMM 
should also be verified for higher orders in $\tilde{g}_s$. 
For example, it will be possible to compute the index in PWMM 
to see if it agrees in the fivebrane limit with the 
index of a six-dimensional (2,0) superconformal field theory 
\cite{Kim:2012qf,Kim:2013nva}, which is supposed to be
the low energy theory of LST.
Also another useful relation to check the existence of 
the NS5-brane limit was proposed 
in \cite{Ling:2006up}. Suppose that the NS5-brane limit exists in PWMM 
and an operator ${\cal O}$ in PWMM has a good scaling law
under the NS5-brane limit, then the coefficients $f_n$ of
the 't Hooft expansion $\langle {\cal O} \rangle =\sum_n f_n/N_2^{2n}$
satisfy
\begin{align}
\frac{f_{n+1}}{f_n}= c \lambda^{5/4} e^{2a \frac{\lambda^{1/4}}{N_5}},
\label{prediction NS5}
\end{align}
where $c$ is a constant. This relation is obtained by 
equating the 't Hooft expansion and the expansion
with respect to $\tilde{g}_s$.
By adding the $1/N_2$ corrections to the analysis of the Fermi gas model 
in the NS5-brane limit, we may be able to answer this problem.

Secondly, though we studied the vacua (\ref{irr decom}) with $\Lambda=1$ 
in this paper, 
the sector of $\phi$ can be mapped to a Fermi gas system
also in the case of $\Lambda \neq 1$. 
In this case the Fermi particles have a labeling, $s=1,\cdots, 
\Lambda$, and the form of the interaction depends on $s$. 
The remaining problem is whether we can see the 
emergent geometry in such a general situation.

Thirdly, the bubbling geometries were also constructed for 
other $SU(2|4)$ symmetric field theories such as ${\cal N}=8$ SYM on 
$R\times S^2$ and ${\cal N}=4$ SYM on $R\times S^3/Z_k$.
The double scaling limits to the NS5-brane solutions were also proposed 
for these theories \cite{Ling:2006xi}.
On the gauge theory side, the localization was also applied to 
these theories \cite{Asano:2012zt}. So we can study 
these cases in the same manner as PWMM.

We hope to report on these issues in the near future.

\section*{Acknowledgements}
We would like to thank H.~Kawai, S.~Kim, S.~Moriyama H.~Shimada 
and T.~Suyama for discussions. 
This work is supported in part by the JSPS Research Fellowship 
for Young Scientists.

\appendix 

\section{Definition of $f^{(n)}_\kappa(x)$}
\label{Definition of f}

$f^{(n)}_\kappa(x)$ is defined as a function satisfying the following Fredholm integral equation of the second kind \cite{Ling:2006up},
\begin{align}
f^{(n)}_\kappa(x)-\int_{-1}^1dyK_\kappa(x,y)f^{(n)}_\kappa(y)=x^n, \label{integral eq of fn}
\end{align}
with kernel
\begin{align}
K_\kappa(x,y)=\frac{1}{\pi}\frac{2\kappa}{4\kappa^2+(x-y)^2}.
\end{align}
For integer $n$, $f^{(n)}_\kappa(-x)=(-1)^{n}f^{(n)}_\kappa(x)$. 
$f^{(0)}_\kappa(x)$ and $f^{(2)}_\kappa(x)$ are relevant for our problem.
In the large-$\kappa$ or the small-$\kappa$ limit, 
one can solve this integral equation.

\subsubsection*{Small-$\kappa$ limit}
For $\kappa\ll 1$, the solution of \eqref{integral eq of fn} can be approximated to
\begin{align}
\frac{1}{2\kappa}\int_{-1}^{1}dy k(x,y)x^n
\label{a3}
\end{align}
with 
\begin{align}
k(x,y)=\frac{1}{2\pi}\log \left\{
\frac{1-xy+\sqrt{(1-x^2)(1-y^2)}}{1-xy-\sqrt{(1-x^2)(1-y^2)}}
\right\}.
\end{align}
For $n=0$ and $n=2$, (\ref{a3}) is evaluated as
\begin{align}
f^{(0)}_\kappa(x)&\simeq \frac{1}{2\kappa}(1-x^2)^{\frac{1}{2}}, \n
f^{(2)}_\kappa(x)&\simeq \frac{1}{2\kappa}\left\{\frac{1}{2}(1-x^2)^{\frac{1}{2}}-\frac{1}{3}(1-x^2)^{\frac{3}{2}}\right\}.
\end{align}
In this case, $\beta(\kappa)$, $q(\kappa)$ and $f_\kappa(x)$
introduced in section \ref{Gravity dual of PWMM} are given by
\begin{align}
\beta(\kappa)&\simeq \kappa, \n
q(\kappa)&\simeq \frac{1}{8}, \n
f_\kappa(x)&\simeq \frac{1}{3\kappa}(1-x^2)^{\frac{3}{2}}.
\label{small k f}
\end{align}

\subsubsection*{Large-$\kappa$ limit}
For $\kappa\gg 1$, one can neglect the second term of the left-hand side of \eqref{integral eq of fn}.
So, for $n=0$ and $n=2$, the leading behavior is
\begin{align}
f^{(0)}_\kappa(x)&\simeq 1, \n
f^{(2)}_\kappa(x)&\simeq x^2.
\end{align}
Hence, $\beta(\kappa)$, $q(\kappa)$ and $f_\kappa(x)$ are approximated to
\begin{align}
\beta(\kappa)&\simeq 2\kappa, \n
q(\kappa)&\simeq \frac{8}{3\pi}\kappa, \n
f_\kappa(x)&\simeq 1-x^2.
\label{large k f}
\end{align}

%%%%%%%%%%%%%%%%%%%%%%%%%%%%%%%%%%%%%%%%%%%%%%%%%%%%%%%%%%%%%%%%%
\section{Off-shell supersymmetries in PWMM}
\label{Off-shell supersymmetry in PWMM}
%%%%%%%%%%%%%%%%%%%%%%%%%%%%%%%%%%%%%%%%%%%%%%%%%%%%%%%%%%%%%%%%%
In this appendix,
we review an off-shell supersymmetries which leave $\phi$ defined 
in (\ref{phi}) invariant \cite{Asano:2012zt}. 
We use the convention in \cite{Asano:2012zt}.
In particular, we work in the Lorentzian signature by making a 
Wick rotation $i X_{10}\rightarrow X_{0}$, so that the bosonic 
symmetry of PWMM is now $R\times SO(3)\times SO(5,1)$. 
See appendix~A in \cite{Asano:2012zt}
for the definitions of the gamma matrices $\Gamma$ and $\tilde{\Gamma}$ used
below.

The supersymmetry transformation in PWMM is given by
\begin{align}
\delta_s X_M&=-i\Psi \Gamma_M \epsilon, \n
\delta_s \Psi&=\left(\frac{1}{2}F_{MN}\Gamma^{MN}
-\frac{1}{2}X_m\tilde{\Gamma}^m\Gamma^a\nabla_a \right)\epsilon,
\end{align}
where the index $a$ runs from $1$ to $4$, 
$\nabla_1 = \frac{\partial}{\partial \tau}$
and $\nabla_{2,3,4}$ are defined as 
\begin{align}
\nabla_{a'} \epsilon = \frac{1}{4}
\varepsilon_{a'b'c'}\Gamma^{b'c'} \epsilon,
\end{align}
The primed indices run from $2$ to $4$.
$\epsilon$ is a real conformal Killing spinor 
with 16 components satisfying 
\begin{align}
\nabla_a \epsilon=\tilde{\Gamma}_a\tilde{\epsilon},
\label{KS eq 1}
\end{align}
where $\tilde{\epsilon}$ is another real spinor satisfying
\begin{align}
\Gamma^a\nabla_a \tilde{\epsilon}&=-\frac{1}{2}\epsilon. 
\label{KS eq 2}
\end{align}
Here $\epsilon$ is Grassmann even, so that $\delta_s$ is Grassmann odd.
One can easily solve these equations 
with the ansatz $\tilde{\epsilon}=\pm\frac{1}{2}\Gamma^{19}\epsilon$, for which
\eqref{KS eq 1} and \eqref{KS eq 2} become
\begin{align}
\nabla_a \epsilon=\pm\frac{1}{2}\Gamma^a\Gamma^{19}\epsilon.
\label{KS eq}
\end{align}
Then, the solution is given by
\begin{align}
\epsilon_+=
\begin{pmatrix}
e^{\frac{\tau}{2}} \ \eta_1 \\
 0 \\
e^{-\frac{\tau}{2}} \ \eta_3 \\
 0 \\
\end{pmatrix} \quad \text{and} \quad
\epsilon_-=
\begin{pmatrix}
 0 \\
e^{-\frac{\tau}{2}} \ \eta_2 \\
 0 \\
e^{\frac{\tau}{2}} \ \eta_4
\end{pmatrix},
\label{KS}
\end{align}
for the upper and the lower sign in (\ref{KS eq}), respectively.
$\eta_{1,2,3,4}$ are four-component constant spinors.
One can see that when the SUSY parameter is given by $\epsilon_+$ with 
$\eta_3=-J_4\eta_1$, $\phi$ is invariant, $\delta_s \phi =0$.
See \cite{Asano:2012zt} for the notation of $J_4$.

By introducing seven auxiliary fields $K_i (i=1,2,\cdots,7)$ 
with the quadratic action,
\begin{align}
\frac{1}{g^2}\int d\tau \frac{1}{2}{\rm Tr}K_iK_i,
\end{align}
one can make the SUSY off-shell \cite{Berkovits:1993hx},
\begin{align}
&\delta _sX_M=-i\Psi \Gamma _M\epsilon ,\nonumber \\
&\delta _s\Psi =\frac{1}{2}F_{MN}\Gamma ^{MN}\epsilon 
-X_m\tilde{\Gamma}^{m}\Gamma^{19}\epsilon 
+K^i\nu _i,\nonumber \\
&\delta _sK_i=i\nu _i\Gamma ^MD_M\Psi.
\end{align}
Here, ${\nu_i}$ are spinors which can be determined by 
the closure of the SUSY algebra.
In particular, when $\epsilon$ is given by $\epsilon_+$ in 
(\ref{KS}) with $\eta_3=-J_4\eta_1$, 
these spinors are explicitly given by
\begin{align}
&\nu _i=\sqrt{2}
e^{\frac{\tau}{2}\Gamma ^{09}}e^{-\frac{\pi}{4}\Gamma ^{49}}\Gamma ^{i8}
\begin{pmatrix}
\eta _1\\
0\\
0\\
0
\end{pmatrix}. \;\;\;\;\; (i=1,2,\cdots,7)
\label{nu}
\end{align}

Since $\{ \Gamma ^{M'}\epsilon,\; \nu ^i|M'=1,\cdots,9,\; i=1,\cdots,7 \}$ 
forms the orthogonal basis of 16 component spinors, 
$\Psi$ can be decomposed as
\begin{align}
\Psi =\Psi _{M'}\Gamma ^{M'}\epsilon +\Upsilon _i\nu ^i.
\label{new basis}
\end{align}
We also define 
\begin{align}
H_i&:=(\epsilon \epsilon )K_i+2\nu _i\tilde \epsilon X_0+s_i,\\
s_i&:=\nu _i\left( \frac{1}{2}\sum_{P,Q=1}^9F_{PQ}\Gamma ^{PQ}\epsilon 
-2\sum_{m=5}^9X_{m}\Gamma ^{ m}\tilde \epsilon \right).
\end{align}
For the SUSY with $\epsilon_+$ with $\eta_3=0$,
by introducing the collective notation,
\begin{align}
X&:=
\begin{pmatrix}
X_{M'}\nonumber \\
(\epsilon \epsilon )\Upsilon _i
\end{pmatrix}, \;\;
X':=
\begin{pmatrix}
-i(\epsilon \epsilon )\Psi _{M'}\\
H_i
\end{pmatrix},
\end{align}
the transformation rules can be written in a compact form as
\begin{align}
&\delta _sX=X',\;\;\;
\delta _sX'=-i(\delta _\phi +\delta _{U(1)})X, \;\;\;
\delta _s\phi =0.
\end{align}
Here, $\delta_{U(1)}$ is a variation under a $U(1)$ subgroup 
of the $SO(3)\times SO(5,1)$ and $\delta_\phi$ is a gauge 
transformation with the parameter $\phi$.
One can see that $\delta_s \Psi_1$ is proportional to 
$\delta_\phi X_1 = D_1 \phi$ as mentioned above (\ref{wi}).

%%%%%%%%%%%%%%%%%%%%%%%%%%%%%%%%%%%%%%%%%%%%%%%%%%%%%%%%%%%%%%%%%
\section{Thomas-Fermi approximation}
\label{Thomas-Fermi approximation}
%%%%%%%%%%%%%%%%%%%%%%%%%%%%%%%%%%%%%%%%%%%%%%%%%%%%%%%%%%%%%%%%%
In this appendix, we review the Thomas-Fermi approximation, which 
is the semi-classical limit of the Hartree approximation.
We consider a one-dimensional many-body system at finite temperature 
$1/\beta$ that has 
a one-body Hamiltonian of the form $h(q,p)=T(p)+U(q)$ and 
a two-body interaction potential $W(q,q')$.
The Hartree approximation is just the saddle-point evaluation of 
the path integral of this system and 
becomes exact when the particle number $N$ goes to infinity.
In this approximation, the saddle point is characterized
by the mean-field density $\rho(x)$ that satisfies the normalization
\begin{align}
\int dq \rho(q) =N.
\label{normalization of rho}
\end{align} 
$\rho(x)$ is determined by the following Hartree equation.
\begin{align}
\rho(q)= \left\langle q
\left| 
\frac{1}{e^{\beta(H(\hat{p},\hat{q})-\mu)}+1}
\right| q
\right\rangle,
\label{Hartree eq}
\end{align}
where $\mu$ is the chemical potential and 
$H(p,q)$ is the effective one-body Hamiltonian defined by
\begin{align}
H(p,q)=T(p)+U(q)+\int dq' W(q,q') \rho(q').
\end{align}
If one obtains $\rho(x)$ by solving the equation (\ref{Hartree eq}),
then from (\ref{normalization of rho}) one can also compute 
the first derivative of the grand potential as 
\begin{align}
\frac{\partial J}{\partial \mu} = \int dq \rho (q).
\end{align}
The free energy is given by 
\begin{align}
F=\log Z=J(\mu(N))-\mu(N) N. \label{F}
\end{align}

In the semi-classical limit, the Hartree equation
(\ref{Hartree eq}) reduces to 
\begin{align}
\rho(q)= \int \frac{dp}{2\pi \hbar} 
\frac{1}{e^{\beta(H(p,q)-\mu)}+1}.
\label{TF eq}
\end{align}
This equation is called the Thomas-Fermi equation at finite temperature.
When the temperature goes to zero, the equation (\ref{TF eq}) is 
further simplified to
\begin{align}
\rho(q)= \int \frac{dp}{2\pi \hbar} \theta(\mu-H(p,q)).
\label{TF zero}
\end{align}
Let us assume that the Fermi surface $\{(p,q)| \mu=H(p,q) \}$
is simply connected and symmetric under $p \rightarrow -p$.
Then (\ref{TF zero}) implies that $\rho(q)$ is given by 
\begin{align}
\rho(q) = \frac{p_F(q)}{\pi \hbar},
\label{rho and pf}
\end{align}
where $p_F(q)$ is the Fermi momentum.
From the definition of $p_F(q)$, we obtain 
the following integral equation that determines $\rho(q)$.
\begin{align}
\mu = T(\pi \hbar \rho(q))+U(q) +\int dq' W(q,q')\rho(q').
\label{integral equation}
\end{align}
This equation can be regarded as an extremization condition for the Thomas-Fermi functional,
\begin{align}
E_{\mathrm{TF}}[\rho]
=\int dq \: t_{\mathrm{TF}}(q)+\int dq \rho(q) U(q) +\frac{1}{2}\int dq dq' \rho(q)W(q,q')\rho(q')
-\mu \left(\int dq \rho(q)-N\right) \label{TF functional}
\end{align}
where $t_{\mathrm{TF}}(q)$ is the kinetic energy functional
\begin{align}
t_{\mathrm{TF}}(q)=\int \frac{dp}{2\pi \hbar} T(p) \theta(\mu - H(p,q)),
\end{align}
and $\mu$ is the Lagrange multiplier associated with the constraint \eqref{normalization of rho},
which can be identified with the chemical potential at the saddle point.
The free energy is given by \eqref{TF functional} with $\rho$ satisfying \eqref{integral equation};
\begin{align}
F=-\mathrm{min}_\rho \: E_{\mathrm{TF}}[\rho]. \label{F2}
\end{align}

%%%%%%%%%%%%%%%%%%%%%%%%%%%%%%%%%%%%%%%%%%%%%%%%%%%%%%%%%%%%%%%%%
\section{Saddle-point method for the D2-brane limit}
\label{Solving the D2 brane limit at the planar level}
%%%%%%%%%%%%%%%%%%%%%%%%%%%%%%%%%%%%%%%%%%%%%%%%%%%%%%%%%%%%%%%%%
In this appendix, we solve our matrix integral for the D2-brane limit in the 
planar limit by applying the usual saddle-point method \cite{Suyama:2011yz}. 
We assume the one-cut solution.
The matrix integral in this limit is given by 
\begin{align}
Z=\int \prod_i dq_i \prod_{i<j} 
\tanh^2 \left(\frac{\pi (q_i-q_j) }{2} \right) 
e^{-\frac{2\pi}{g_{S^2}^2}\sum_i q_i^2}.
\end{align}
By changing the integral variables to $z_i:=\exp (\pi q_i +g^2_{S^2}\pi/4) $,
the path integral is reduced to 
\begin{align}
Z= \int \prod_i dz_i \prod_{i>j}
\left(\frac{z_i-z_j}{z_i+z_j} \right)^2 
e^{-\frac{2}{g_{S^2}^2\pi}\sum_i (\log z_i)^2}.
\end{align}
The saddle-point equation is given by 
\begin{align}
\frac{2}{g^2_{S^2} \pi} \frac{\log z_i}{z_i}
- \sum_{j(\neq i)}\left(\frac{1}{z_i-z_j}-\frac{1}{z_i+z_j} \right)=0.
\label{saddle point}
\end{align}
Note that this equation is symmetric under the inversion, 
$z_i \rightarrow 1/z_i$.
Let $[a,b]$ be the support of
the eigenvalue distribution of $z_i$.
From the inversion symmetry, it follows that $b=1/a$.
We define the resolvent as
\begin{align}
W(z)= 4g^2_{S^2} \pi \sum_i \left(\frac{1}{z-z_i}-\frac{1}{z+z_i} \right).
\end{align}
This function has two branch cuts at $[a,b]$ and $[-b,-a]$.
The eigenvalue distribution,
\begin{align}
\rho(z)= \frac{1}{N_2}\sum_i \delta (z-z_i),
\end{align}
can be expressed as the discontinuity of $W(z)$ as usual,
\begin{align}
W(z+i0)-W(z-i0)= -8\pi^2 i g^2_{S^2} N_2 \rho(z).
\label{W-W}
\end{align}

We introduce a new variable $y=z^2$. 
The resolvent is also a holomorphic function of $y$. 
So let us denote $W(z)=P(y)$, where $P(y)$ is holomorphic in $y$. 
$P(y)$ has a single cut at $[a^2,b^2]$ on the $y$-plane.
Using (\ref{saddle point}), one can easily get 
\begin{align}
P(y+i0)+P(y-i0)=\frac{8\log y}{\sqrt{y}},
\label{P+P}
\end{align}
where $y \in [a^2,b^2]$. 
By defining a new function,
\begin{align}
\hat{P}(y)= \frac{P(y)}{\sqrt{(y-a^2)(y-b^2)}},
\end{align}
one can convert (\ref{P+P}) to the discontinuity equation,
\begin{align}
\hat{P}(y+i0)-\hat{P}(y-i0)=\frac{1}{\sqrt{(y-a^2)(y-b^2)}}
\frac{8\log y}{\sqrt{y}}.
\end{align}
This equation determines $\hat{P}$ up to the regular part. 
Since $\hat{P}(y)\sim y^{-2}$ when $y \rightarrow \infty$, 
the regular part should be vanishing. 
Thus, we obtain 
\begin{align}
\hat{P}(y)=\int_{a^2}^{b^2} \frac{dp}{2\pi} \frac{8 \log p}{(y-p)\sqrt{p}}
\frac{1}{\sqrt{(p-a^2)(b^2-p)}},
\end{align}
and then the resolvent is given by
\begin{align}
W(z)= 32 \int_a^b \frac{dq}{2\pi} \frac{\log q}{z^2-q^2}
\sqrt{\frac{(z^2-b^2)(z^2-a^2)}{(b^2-q^2)(q^2-a^2)}}.
\label{W int}
\end{align}
From (\ref{W-W}), the eigenvalue distribution is given by
\begin{align}
\rho(x)=\frac{4}{\pi^3 g^2_{S^2}N_2} P\int_a^b dq
\frac{\log q}{q^2-x^2}
\sqrt{\frac{(b^2-x^2)(x^2-a^2)}{(b^2-q^2)(q^2-a^2)}},
\label{rho int}
\end{align}
where $x \in [a,b]$ and $P\int$ means the principal value.
Note that it satisfies
\begin{align}
x\rho(x)=\frac{1}{x}\rho(1/x).
\label{rho inversion}
\end{align}

When the 't Hooft coupling $g^2_{S^2}N_2$ is large, the integral 
in (\ref{rho int}) can be performed. This limit will turn out to 
correspond to the large-$b$ limit.
By changing the variables in (\ref{rho int}) as
\begin{align}
\log b= -\log a= \alpha, \;\;\;  \log x =v \alpha, \;\;\; \log q= u \alpha,
\end{align}
one can obtain
\begin{align}
x\rho(x)+\frac{1}{x}\rho(1/x)= \frac{4\alpha^2}{\pi^3 g^2_{S^2} N_2}
P\int_{-1}^1 du \; u \;{\rm sign}(u-v) + {\cal O}(\alpha^1).
\end{align}
Then, the integral can be easily performed. By using 
(\ref{rho inversion}), one obtains
\begin{align}
\rho(x)= \frac{2}{\pi^3 g^2_{S^2}N_2} \frac{(\log b)^2-(\log x)^2}{x},
\label{rho solution}
\end{align}
in the leading order of $\alpha$.
Since $\int_a^b \rho(x)dx =1$ by definition, $b$ is determined as
\begin{align}
b=1/a=\exp \left[
\pi \left(\frac{3g^2_{S^2}N_2}{8} \right)^{\frac{1}{3}}
\right].
\end{align}
Thus, $b$ is indeed large when the 't Hooft coupling is large.

Using (\ref{rho solution}), one can easily compute the free energy 
of the matrix integral. The result is given by
\begin{align}
\log Z = \frac{9\pi }{10} \frac{N^2_2}{(3g_{S^2}^2N_2)^{1/3}}.
\label{free energy on s2}
\end{align}
%One may naively think that this is the free energy of 
%${\cal N}=8$ SYM on $R\times S^2$ around the trivial vacuum.
%However, this is not the case because the localization 
%calculation breaks some of supersymmetries and also the 
%translational invariance of PWMM. 
%As a result, (\ref{free energy on s2}) is finite 
%and non-zero. On the other hand,
%the free energy of ${\cal N}=8$ SYM on $R\times S^2$
%should be proportional to the volume of the space-time
%and should be vanishing for a theory around 
%any supersymmetric vacuum.
%So one should regard (\ref{free energy on s2}) just as
%a generating function of ${\rm Tr}M^2$ not the free energy
%of the gauge theory.

\end{document}